# Spatiotemporal mode-locking in multimode fiber lasers

Logan G. Wright[1]*, Demetrios N. Christodoulides[2], and Frank W. Wise[1]

[1]School of Applied and Engineering Physics, Cornell University, Ithaca, New York 14853, USA

[2]CREOL, College of Optics and Photonics, University of Central Florida, Orlando, Florida 32816, USA

*Correspondence to: lgw32@cornell.edu

**Abstract**: A laser is based on the electromagnetic modes of its resonator, which provides the feedback required for oscillation. Enormous progress has been made in controlling the interactions of longitudinal modes in lasers with a single transverse mode. For example, the field of ultrafast science has been built on lasers that lock many longitudinal modes together to form ultrashort light pulses. However, coherent superposition of longitudinal and transverse modes in a laser has received little attention. We show that modal and chromatic dispersions in fiber lasers can be counteracted by strong spatial and spectral filtering. This allows locking of multiple transverse and longitudinal modes to create ultrashort pulses with a variety of spatiotemporal profiles. Multimode fiber lasers thus open new directions in studies of nonlinear wave propagation and capabilities for applications.

The modes of a laser resonator are three-dimensional (3D) functions that vary along the axis of the resonator as well as in the two transverse dimensions *(1)* (Fig. 1). Each mode has a distinct resonant frequency (Fig. 1B, D). In many cases of interest, the 3D modes are separable into so-called longitudinal and transverse modes. If the relative phases of the modes are not controlled, the output is an incoherent spatiotemporal field that results from random interference. The modes can interact with each other in the gain medium and other components of a laser.

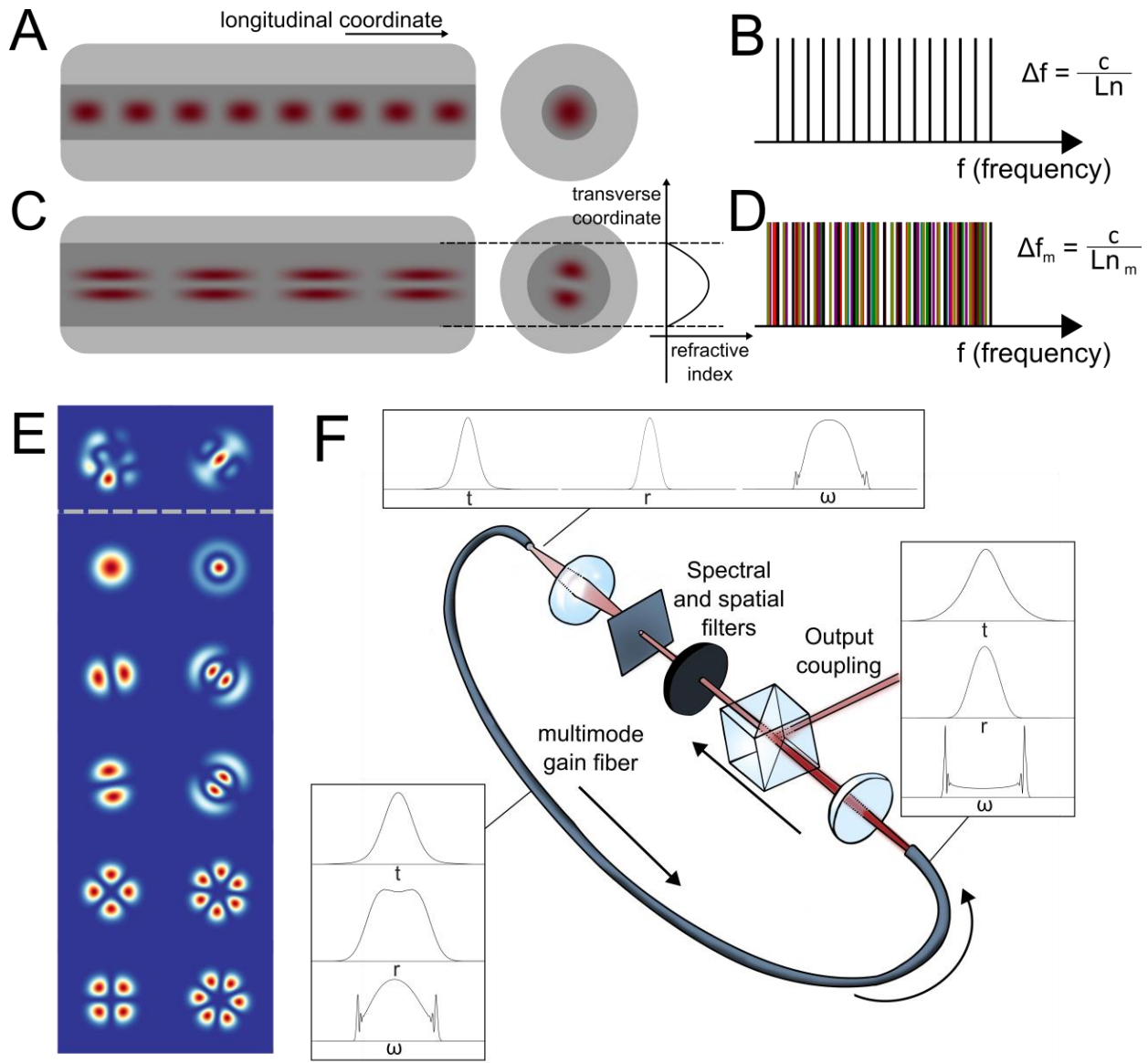

**Figure 1: Conceptual outline of mode-locked multimode fiber lasers.** With a single transverse mode, lasing occurs in longitudinal eigenmodes, which correspond to different patterns of the electromagnetic lasing field along the length of the fiber (**A**), and resonant frequencies (**B**). In a multimode fiber, the resonant frequency spectrum (**D**) and transverse field distribution (two examples shown at the top of **E**) can be complex. A family of longitudinal modes corresponds to each transverse mode (lower part of **E**, different colors in **D**). In a multimode fiber, all possible field distributions can be decomposed into contributions from these transverse eigenmodes (**E**). In a multimode fiber laser cavity, the lasing eigenmodes describe 3D electromagnetic field patterns. One 3D mode is illustrated in **C**. **F.** Simulated evolution of intensity profiles in a spatiotemporally-mode-locked state.

The situation can be simplified greatly by restricting operation to a single transverse mode. With the exception of high average power, the best performance for virtually all laser parameters -- ultranarrow emission spectra, ultrashort pulse duration, ultralow noise, e.g. -- is achieved through lasing in a single transverse mode. Operation in multiple spatial modes, by contrast, leads to significant complexity. It is nonetheless widely employed in high-average-power lasers, for applications that do not require high temporal or frequency precision, nor diffraction-limited focusing.

Highly-refined techniques have been developed to control the longitudinal modes of a laser. The number of oscillating modes can be as few as one, to more than one million perfectly-synchronized modes in a frequency comb *(2)*. Considering the exquisite control of the electromagnetic field achieved in lasers with a single transverse mode, the lack of attention paid to higher-dimensional coherent lasing states is conspicuous. Early works did propose locking of transverse spatial modes, as well as transverse and longitudinal modes. Mode-locking of this kind was demonstrated with a few modes *(3,4)*, and has since been observed with two transverse modes of a femtosecond Ti:sapphire laser *(5)*.

Recent years have seen increased interest in linear and nonlinear wave propagation in multiple transverse modes, mainly in multimode optical fibers *(6-16)*, and largely in anticipation of spatial division multiplexing *(17)* for telecommunications and as a platform for new fiber laser sources. The presence of disorder (through random linear mode coupling) also connects multimode fiber to the field of random lasers, and more broadly to the optical properties of complex media *(18-21)*.

Here we employ principles of normal-dispersion mode-locking *(22-23)* in space and time – strong spectral and spatial filtering in addition to the high nonlinearity, gain and spatiotemporal dispersion of the fiber medium – to achieve spatiotemporal mode-locking. The self-organized, mode-locked pulses take a variety of spatiotemporal shapes consisting of many spatial and longitudinal modes.

Lasing modes can interact through the electronic nonlinearities of the gain, the saturable absorber, and the fiber medium itself. These effects occur on the time scale of a pulse, and thus can couple temporal and spatial degrees of freedom. The slow relaxation of rare-earth gain media introduces an additional layer of temporally-averaging nonlinear interactions. To realize the simplest demonstration of highly-multimode spatiotemporal mode-locking (STML), we began by constructing a cavity with few-mode, Yb-doped gain fiber (10 μm diameter, supporting ~3 transverse modes) spliced to a highly-multimode (MM) passive graded-index (GRIN) fiber (which supports ~100 transverse modes), thereby minimizing transverse gain interactions and isolating the key nonlinear interactions involved in passive mode-locking. GRIN fiber is used so that the modal dispersion within the cavity is relatively small, comparable to the chromatic dispersion. As will be expanded on later, this is an important factor for achieving STML. Excitation of many modes of the GRIN fiber is accomplished by fusing the two fibers with varying spatial offsets. We employ nonlinear polarization rotation as an ultrafast saturable absorber. This cavity is readily-modelled by a set of coupled nonlinear Schrödinger equations (NLSEs), and has the secondary benefit of allowing observation of highly-MM mode-locking at relatively low laser powers. As a result, the cavity serves as a convenient experimental and theoretical test-bed for multimode laser dynamics. Simulations reveal a rich variety of stable spatiotemporal pulses (an example is shown in Fig. 2A-C, and Fig. S1-4, while others are shown in Fig. S5-12). These pulses form from noise and take on a variety of different modal compositions.

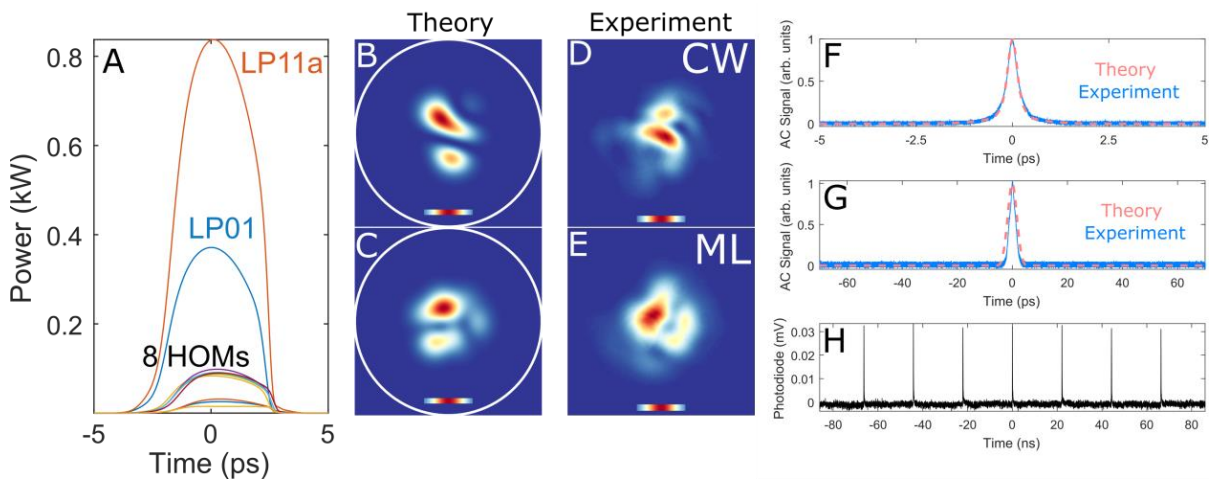

**Figure 2: Characterization of a spatiotemporally-mode-locked pulse train, theory and experiment. A** shows the simulated mode-resolved pulse intensity in 10 transverse modes. **B** and **C** show the simulated continuous wave (CW) and mode-locked near-field beam profiles, while **D** and **E** show the same for the experiment. **F** shows the dechirped

intensity autocorrelation for both simulation and experiment, while **G** shows the autocorrelation for the chirped pulse over a 140 ps range. **H** shows the pulse train measured using a photodiode with a ~40 ps resolution. **A-H** show that the multimode beam in **E** corresponds to a single pulse that comprises >10 locked transverse modes, with no continuous-wave background. This figure is extended in Fig. S1-4. Real-time acquisition of the experimental measurements, additional demonstrations, and animations of the numerical results, are shown in **Movie S1**. All mode-locked states examined in this manuscript are self-starting.

The straightforward emergence of stable spatiotemporal mode-locking is surprising, but can be understood through the comparable dispersion of transverse and longitudinal modes in GRIN fiber-based cavities, and the periodic spatial and spectral filtering we employ in the cavity. The formation and stability of these 3D mode-locked pulses is conceptually similar to 1D dissipative solitons and self-similar pulses in single-mode fiber (*22, 23*). In these lasers, normal dispersion and nonlinear phase modulation lead to a chirped pulse whose duration and bandwidth grows through most of the cavity. Strong spectral filtering of this chirped pulse reduces the duration and bandwidth, and so allows the pulse to meet the periodic boundary condition of the laser. In the current study, this approach is applied in space-time: in the multimode cavities, the combination of spatial and spectral filtering helps to establish a 3D steady-state pulse evolution with periodic boundary conditions in both space and time (Fig. S3-4), a crucial aspect of mode-locking. Since in GRIN fiber the magnitude of transverse mode dispersion is similar to the chromatic dispersion, coupling between all types of modes is equally strong. Hence in general mode-locking in such a 3D cavity involves numerous kinds of 3D modes.

Experimentally, spatial and spectral filtering are implemented through the overlap of the MM field with the gain fiber input, and a bandpass interference filter (Fig. S13). With suitable adjustments of the filters and wave-plates (an experimental algorithm is described in the methods), a variety of highly MM (10-100 transverse modes) spatiotemporal mode-locking states can be observed (Fig. 2D-H, Fig. S14-16). The typical pulse energies range from 5 to 40 nJ, corresponding to routinely-available pump powers, and peak powers well below the threshold for end-facet damage. The clearest evidence for spatiotemporal mode-locking is the sudden transition in the spatial, spectral and temporal properties as we increase or decrease the pump power through the mode-locking threshold. Since we have eliminated laser gain interactions in this cavity, our observations of a different beam profile when mode-locking occurs at slightly higher pump power can only be explained by a spatiotemporal mode-locking transition involving nonlinear interactions between many transverse families of longitudinal modes. Without gain interactions, and because the output is taken directly after the passive GRIN fiber, the CW spatial beam profile also reveals the active lasing modes in the cavity when the laser is mode-locked. Temporal measurements (such as shown in Fig. 2F-G, Fig. S1-2 and Movie S1) demonstrate that this spatiotemporal self-organization results in single pulses comprising many non-degenerate transverse mode families, as predicted by our simulations.

For understanding and applications alike, it is important to determine if STML can still occur when slow transverse gain interactions are present. Informed by results from the first cavity, we investigated spatiotemporal mode-locking in a second cavity based completely on a partially-graded, highly multimode Yb-doped gain fiber (Fig. S17). With necessarily weaker spatial filtering, and richer mixture of spatial and spatiotemporal nonlinear interactions, theoretical modelling and analysis of this scenario is more complicated than the previous cavity. Nonetheless, we developed a qualitative spatiotemporal model that incorporates all relevant effects. This model (Supplementary Text) shows that the key principals of STML carry over from the simpler cavity. With high power, and spectral and spatial filtering, we observe many different complex spatiotemporal dynamics in numerical simulations (Figs. S18-25), including stable mode-locked pulse trains.

We constructed a corresponding laser similar to the previously-described cavity, except all of the fiber is the highly-multimode, partially-GRIN Yb-doped gain fiber, and spatial filtering is supplemented by an adjustable slit or iris (Fig. 1F, methods, Fig. S26). Above a threshold pump power, STML lasing states are observed (Fig. 3). As the pump power is adjusted, we observe discontinuous changes in the field's spatial, temporal and spectral properties as the laser undergoes the transition from CW lasing in 10-100 transverse modes to multimode mode-locking where the output beam comprises ~2-10 modes. As with the previous cavity, this observation provides strong evidence for STML. Further temporal measurements (Fig. 3D, E) and additional analysis provide strong evidence that the produced pulse comprise multiple transverse mode families (for further details see methods, Supplementary Text).

For this cavity, again a wide variety of different mode-locked states are observed for different cavity configurations (Fig. S29). However, the minimum power for which mode-locking occurs in this laser corresponds to intracavity powers >100 kW. Consequently, systematic exploration of this laser and future engineering of practical instruments will require application of known techniques for avoiding fiber damage. Furthermore, direct measurements of the 3D field will be an important step for increased understanding of STML. Ongoing developments in this direction are promising (e.g., Ref. 24).

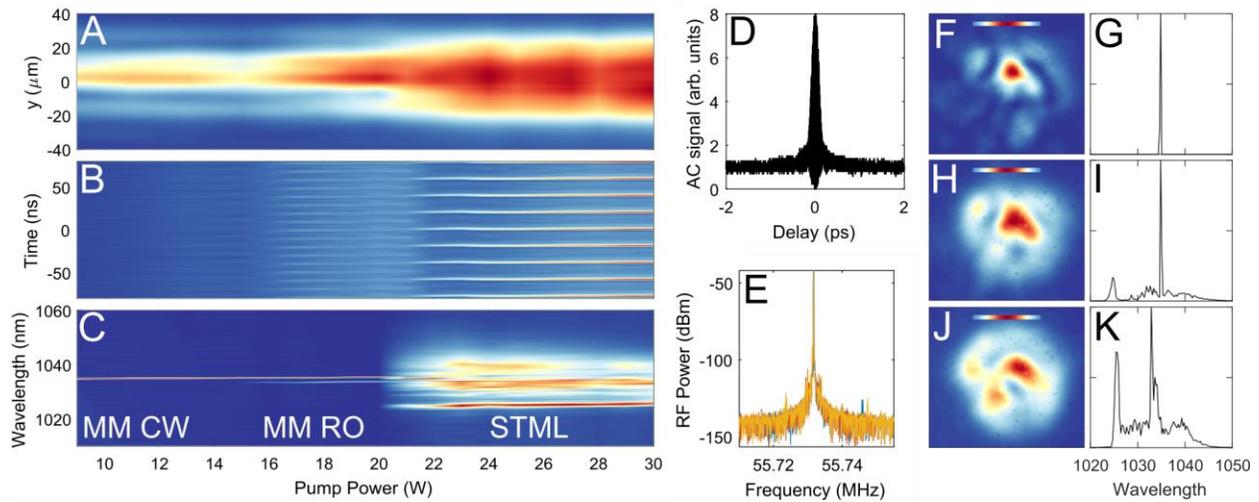

**Figure 3: Transition to mode-locking in the second laser cavity.** Variation with pump power of (**A**) near-field beam profile integrated over 1 dimension, (**B**) temporal variation of the output, and (**C**) spectrum. As the pump power is changed, the field transitions from multimode continuous-wave lasing (MM CW, **F** and **G**) to relaxation oscillations (MM RO), to bistability with relaxation oscillations (~20-23 W, **H** and **I**), and then to STML (**D, E, J, K**). The CW lasing threshold is 9 W. The coherence of the mode-locked state is evident from the autocorrelation of the dechirped pulse (**D**) and the RF spectra (**E**, where the different colors correspond to measurements taken at different, arbitrary positions on the beam). The scale bar in **F, H, J** shows the Gaussian profile of the fundamental mode of the fiber.

Investigation of 3D mode-locking will connect scientifically to areas well beyond short-pulse generation. The dimensionality and controllable dissipation and disorder make consideration of wave turbulence and condensation *(25-26)* especially interesting. With longer or suitably-designed fibers, disorder can be introduced controllably, thereby bridging the gap to random lasers and allowing exploration of the entire parameter space of multimode lasing *(19-20)*. Although disorder is weak in the present work, recent research indicates that multimode self-organization may be robust to disorder *(12-16)*. This will be important for applications, which will ultimately require environmental stability. In addition, mode-locked lasers are closely related to coherently-driven microresonators, where the influence of higher-order transverse modes is starting to be considered (27-29). Spatiotemporal mode-locking and related phenomena should be possible in these systems, and provide new features for frequency comb applications.

The potential of MM mode-locked lasers for high performance is significant. Initial results obtained with the second cavity (150-nJ and 150-fs pulses, for ~1-MW peak power and ~10-W average power) already rival the best achieved with flexible, large-area single-mode fibers *(30)*. With larger fiber core areas, we expect scaling of pulse energy by over 2 orders of magnitude to be possible in principle (Fig. S20-21). At the highest pulse energies, simulations predict the emergence of a stable multimode Gaussian output beam, which will be useful for applications (Fig. S18).

Exploration of the full scope of MM mode-locking will require developments on several fronts, including measurement capabilities. The ability to generate high-power and spatiotemporally-engineered coherent light fields should lead to breakthroughs in laser science as well as applications.

**Acknowledgments:**

L.G.W. performed experiments and simulations. L.G.W. and F.W.W. wrote the first drafts of the manuscript. F.W.W and D.N.C provided funding and supervised the project. All authors contributed to the final manuscript. LGW and FWW have submitted a patent application for spatiotemporally mode-locked lasers. This work was supported by Office of Naval Research grant N00014-13-1-0649 and National Sciences Foundation grant ECCS-1609129. We thank Nufern for providing the partially-graded multimode gain fiber, W. Fu for pointing this product out, Z. Ziegler for discussions and contributions to some of the numerical codes used for MMGNLSE simulations, Z. Liu for discussions and his early contribution to the work, E. Falco for the laser schematic in Fig. 1F, and W. Renninger for his critical review of an early version of the manuscript. All data presented in this report and the Supplementary Material is available on reasonable request to the L.G.W.

# Supplementary Materials for

## Spatiotemporal Mode-Locking in Multimode Fiber Lasers

Logan G. Wright, Demetrios N. Christodoulides, and Frank W. Wise

correspondence to: lgw32@cornell.edu

**This PDF file includes:**

    Materials and Methods
    Supplementary Text
    Figs. S1 to S33
    Captions for Movie S1

**Other Supplementary Materials for this manuscript includes the following:**

    Movies S1

**Materials and Methods**

Simulations – first cavity

Simulations of this cavity are performed using the generalized multimode nonlinear Schrödinger equation (*6,11*). Different combinations of modes are considered, and the dispersions and mode-coupling tensors are calculated from the numerically-calculated modes for a GRIN fiber based on the one used experimentally, with a 50 μm diameter, refractive index power of 2.08, and numerical aperture of 0.2. The gain fiber is approximated as a single-mode fiber. The effective saturable absorber provided by nonlinear polarization rotation is modelled by an idealized, instantaneous saturable absorber transfer function applied to the composite single-mode field to be coupled into the single-mode fiber. Overlap between the gain and multimode fiber is controlled by two sets of mode coupling coefficients describing the excitation of modes at the splice point, along with the overlap of the MM output field with the gain fiber. The gain spectrum is modelled as a Gaussian with 40-nm bandwidth; the small-signal gain is 30 dB. The dispersion of the gain fiber is assumed to be 20 $fs^2$/mm, while for the MM fiber the first 80 calculated transverse modes have normal dispersion approximately 20 $fs^2$/mm. Stable pulses are observed for a wide variety of parameters; more specific details on the model, and representative examples are provided in the supplementary information (see 'Additional simulation details and exemplary results for the first cavity', Figs. S2-S12).

Experiments – first cavity

The first cavity (Fig. S13, Movie S1) was designed to minimize the impact of transverse mode interactions through gain saturation (leaving only the ultrafast interactions responsible for passive mode-locking), and to support strongly spatially multimode mode-locking at low power levels. 1-2 m of 10-μm diameter step-index, few-mode gain fiber (Thorlabs YB1200-10/125DC) is spliced to 0.5-3 m of Corning OM4 graded-index multimode fiber. To excite many modes, the GRIN fiber is aligned offset from the gain fiber's center by up to 25 μm. To minimize reflections from this splice point, the arc fusion splicer is continuously arced until the fibers' claddings are overlapped to a high degree. This process strongly distorts several millimeters of GRIN fiber nearest the splice, ensuring that both highly-multimode excitation and low-loss coupling from the gain fiber into the multimode GRIN fiber occurs. Pump light leakage is controlled by application of index-matching gel and protection by darkened foil. In both cavities, bulk-optic filters, wave-plates, isolators, and beamsplitters are employed in a ~ 1-m long free-space section used for experimental convenience and power handling. In both cavities, the fiber is laid loosely with an approximate coiling diameter of 50 cm, and the fiber is pumped by a multimode, 30-W, 976-nm diode laser (Gauss Laser) through dichroic mirrors.

In order to achieve mode-locking in any cavity, intracavity back-reflections that lead to parasitic lasing must be suppressed. In a multimode cavity, this is particularly important since there are numerous paths for back-reflections to form and travel. To avoid this, careful optimization of the cleave parameters for the angle-cleaved fibers is performed. Similar optimization is required for splices, with MM-exciting splices performed as described above.

With the above measures, STML is easily observed. After aligning the cavity to maximize the CW output power, mode-locking in the cavity is straightforwardly initiated, and different states are accessed by adjusting the waveplates and pump power, and by tuning the mirrors in the cavity to change the overlap between the multimode field that exits the GRIN fiber and the gain fiber. A wide variety of spectral filters and mirror alignments result in many different spatiotemporal pulses. However, due in large part to the placement of a long passive fiber after the gain, we find that at the highest powers strong Raman-based effects lead to pulse distortions and Q-switching. These can be avoided in part by use of filter with a single pass-band (commonly used birefringent filters support many passbands), but ultimately even with this measure, these processes limit the maximum energy of stable STML pulses observed in this cavity.

Experiments – second cavity

The multimode gain fiber cavity (Fig. S26) employs multimode active fiber with a partially-graded index profile (Nufern FUD-7005) shown in Fig. S17. The center 70 µm diameter section of the fiber is doped with Yb. This fiber supports roughly 90 spatial (180 total) modes at 1030 nm, and the graded-index design leads to small modal dispersion compared to step-index fibers. The first 23 calculated modes of this fiber have normal dispersion. Since these modes are the ones most localized to the doped region, and have the smallest modal dispersion, we expect modes with anomalous dispersion do not play any role in the mode-locking observed, although they may become involved in the high-power simulations. This dominance of lower-order modes is consistent with the experimentally-observed mode-locked beam profiles. In order to maximize the nonlinear phase shift and achieve spatiotemporal mode-locking at modest pump powers, 3.5-4.0 m of this fiber are used – significantly more than would be required for an efficient laser cavity. With higher pump powers and/or additional passive fiber, we expect this would not be necessary.

The cavity is initially aligned to maximize continuous wave lasing power without additional spectral or spatial filtering. The finite gain bandwidth and the field's overlap with the isolator aperture and the fiber input provide some spectral and spatial filtering, but for the powers considered here only unstable mode-locking can be achieved without stronger filtering. A single pass-band spectral filter (typically a simple interference filter, Thorlabs FLH1030-10) is then added and a spatial filter (an adjustable slit or pinhole) is closed until the intracavity field is noticeably clipped (the maximum power is generally reduced by 20%). Mode-locking can then be initiated and adjusted by rotating the waveplates. This procedure is generally enough to allow many mode-locked states, and provided the pump power is high enough, mode-locking is straightforward to observe. However, if mode-locking cannot be initiated with this method, the waveplates are adjusted until relaxation oscillations and/or self-Q-switching are evident from measurements of the spectrum or temporal profile with a fast photodetector. At this point, the spatial filter and folding mirrors can be adjusted slightly until mode-locking is observed.

Simulations for this cavity were performed with a qualitative (2+1)-D beam propagation model, which incorporates full spatial gain saturation. This model is described in 'Additional simulation details and exemplary results for the second cavity', and example results are shown in Fig. S18–25.

Experiments – field measurements

The average spectrum of the field exiting the multimode fiber is measured, after sampling by beamsplitters, by coupling it into a long highly multimode fiber connected to an optical spectrum analyzer. The long multimode fiber allows sampling of most of the field, and the long diffusive linear propagation serves to spatially-average the spectrum. In some cases, we observe a spatially-sampled spectrum by coupling only part of the input beam into the spectrum analyzer, using either a single-mode fiber or using the multimode fiber without a focusing lens. When the spectrum of the mode-locked pulse is measured in this way, we observe changes in the spectrum as the beam is translated, which confirms the multimode spatiotemporal structure implied by spatial measurements and numerics (See Figs. S1-2, and Movie S1). The beam profile is measured using a 4-f telescope, where the first lens is aligned by imaging amplified spontaneous emission from the laser cavity onto the camera. To additionally verify that the mode-locked pulses contain multiple transverse modes, we measured the beam profile through a variably-rotated narrow bandpass interference filter (See Fig. S1-2, and Movie S1). We measure the autocorrelation of pulses out of the cavity (which are chirped), and after dispersion compensation by a grating compressor (dechirped). Use of two-photon absorption ensures that the entire spatiotemporal field is integrated into the measurement, without complex phase-matching (*11*). The coherence of the pulses is confirmed by the fact that they can be dechirped to near their spectrum's transform limit, with an 8:1 peak:background ratio in their interferometric autocorrelation. To measure delays over roughly 20 ps, we utilized a long-range (>185 ps) intensity autocorrelator (Femtochrome 103XL) based on second-harmonic generation in a thin crystal. In this case, slight changes in the autocorrelation shape (both for the chirped and dechirped pulse) are observed for different alignments into the autocorrelator. This is consistent with simulations (see 'Discussion of spatiotemporal dispersion in spatiotemporally mode-locked pulses', Figs. S32-33). To facilitate direct comparison with simulations, we compare the alignment producing maximum signal to the numerical autocorrelation over the most intense region of the numerical dechirped pulse. Together with the measurements using the fast (~40 ps resolution) photodiode/oscilloscope, the long-range autocorrelation provides a systematic proof of single-pulse operation. For both autocorrelators, deviations from transform limit in the first cavity are sometimes observed. These deviations can be attributed to Raman scattering, which appears on the red side of the pulse spectrum for high pulse energies, and to spatiotemporal dispersion (see 'Discussion of spatiotemporal dispersion in spatiotemporally mode-locked pulses', Figs. S32-33). The fast photodiode and oscilloscope were used to check for single-pulse mode-locking (Fig. 2H, Fig. S1), but for routine measurements, including when varying pump power or mode-locking, a slower photodiode was used since fast changes in the spatiotemporal field were found to easily cause damage to the more sensitive fast device. In the second cavity, we observed that over several minutes, a stable mode-locked state could destabilize and return very quickly. We attribute this behavior to the proximity of the mode-locking threshold to our maximum pump power (the laser is likely still within the bistable region). Consequently, the fast photodiode was not used for this laser. The slow photodiode's decay results in the appearance of a finite zero-offset during mode-locked operation, possibly implying a CW background (e.g., Fig. 3B, S27-28). To help rule out multi-pulsing or a CW background, radio-frequency (RF) spectral measurements were taken (with a resolution bandwidth of 30 Hz) at several spatial locations (different traces in Fig. 3E). Some features in this spectrum distant from the fundamental correspond to short losses of mode-locking during the measurement time of ~minutes. These measurements show that when the laser is mode-locked, there is a >95 dB contrast between the mode-locked repetition frequency and the noise background. By comparison with measurements of single-mode oscillators, this value strongly suggests that the laser operates in a CW-free, single-pulsing state. Assessment of the peak power is significantly complicated in general by the complex spatiotemporal structure of the pulses. However, for relatively simple output beams, spectral broadening of sampled portions of the beam in single-mode fiber agree with numerical simulations, and the observed single-pulse energy is consistent with the nonlinear phase shift inferred from the spectral breathing of the pulse in the cavity (for further details, see Figs. S30-31 and 'Peak power assessment').

To measure the spatiotemporal mode-locking transition, we choose to decrease the pump power starting from the highest value. After measurements are made for decreasing pump power, the power is returned to the original level. For all results shown here and in the Supplementary Material, nearly identical mode-locking is recovered. In other words, the mode-locked states observed here are self-starting. However, for increasing pump powers we generally observe poorer stability of the mode-locked state. In the all-multimode cavity, peak-power-related end facet damage is regularly observed during the instability of mode-locked states. To avoid fiber damage we therefore performed most detailed measurements with decreasing pump power in experiments.

**Supplementary Text**
Additional simulation details and exemplary results for the first cavity
We performed numerical simulations of the first multimode cavity using the generalized multimode nonlinear Schrödinger equation (GMMNLSE) to describe evolution on the modal field envelopes, $A_p(z,t)$, in the passive multimode GRIN fiber:

$$\partial_z A_p(z,t) = i\left(\beta_0^{(p)} - \Re\left[\beta_0^{(0)}\right]\right)A_p - \left(\beta_1^{(p)} - \Re\left[\beta_1^{(0)}\right]\right)\frac{\partial A_p}{\partial t} + \sum_{m=2}^{3} i^{m+1}\frac{\beta_m}{m!}\partial_t^m A_p + i\frac{n_2\omega_o}{c}\left(1 + \frac{i}{\omega_o}\partial_t\right)\sum_{l,m,n}\{(1-f_R)S_{plmn}^k A_l A_m A_n^* + f_R A_l S_{plmn}^R \int_{-\infty}^{t} d\tau A_m(z,t-\tau)A_n^*(z,t-\tau)h_R(\tau)\}$$

The parameters are defined in Ref. 11. For the examples shown here, the spatial modes are numerically calculated assuming linear polarization at a wavelength of 1030 nm, and the dispersion parameters are obtained by polynomial fits of the dispersion curves obtained by numerically solving for the modes from ~1550 nm to ~ 750 nm. We find that the first 80 guided modes have normal dispersion of roughly 20 fs$^2$/mm. We approximate the propagation in the few-mode gain fiber by a single mode GNLSE in the fundamental mode of that fiber.

At the interface between the gain fiber and the multimode (MM) fiber, and between the MM fiber and the gain fiber, we approximate the MM excitation and the spatial filtering process by projection coefficients, $P_{1p}$ and $P_{2p}$, where $p$ is the index of the mode, running from 1 to 10 for the simulations shown here. That is, the field entering the MM fiber is:
$$A_p(z=0,t) = P_{1p}A_{SMF}(z=L_{SMF},t)$$
The field that re-enters the gain fiber is
$$A_{SMF}(z=0,t) = \sum_{p=1}^{N} P_{2p}A_p(z=L_{MMF},t)$$
In both cases, the field is re-normalized so that the projection conserves energy. After output from the MM fiber, the transformation above is applied and an ideal Gaussian spectral filter is applied. Then a saturable absorber transfer function is applied to the field:

$$A(z,t) \to A(z,t)\sqrt{1 - \frac{mod\_depth}{1 + \frac{|A(z,t)|^2}{I_{sat}}}}$$

This transfer function implicitly assumes an instantaneous saturable absorber, which is a good approximation to the Kerr-nonlinearity derived effective absorber used in our laser (nonlinear polarization rotation). A portion of the field is taken as the output, and an additional lumped loss (50% for all the examples considered here) is applied before the field is propagated through the gain fiber again. The lumped loss accounts for the imperfect fiber coupling, and losses within the spectral filter and isolator. The application of the saturable absorber to the projected single-mode field is a significant simplification over the spatiotemporal saturable absorption which occurs in the experiments (although effective saturable absorption arising through transverse mode interactions, generalized Kerr lensing, is still captured). A subsequent section (Section 4) details simulations using spatiotemporal saturable absorption and multimode gain fiber.

The corresponding supplementary figures show representative examples of multimode mode-locking. An extremely wide range of pulses and behaviors are observed; the included examples are just meant to complement the primary example Fig. 2, Fig. S1-4, in showing the most common features of STML. All simulations were seeded with random noise along with a pulse with amplitude 30% higher than the noise and placed at -15 ps, to encourage formation of a pulse at that point in the temporal window of the simulation. Figs. S5-S8 shows an example where the fundamental mode is the highest-power spatial mode. Figs. S9-S12 show the possibility of multimode mode-locking with many modes having similar amplitudes.

Additional simulation details and exemplary results for the second cavity

We use a (2+1)-D GNLSE to provide a qualitative model for spatiotemporal mode-locking in a cavity based entirely on MM gain fiber. The evolution of the (2+1)-D spatiotemporal pulse envelope, $A(x,z,t)$, is found by solving:

$$\partial_z A(x,z,t) = \frac{i}{2k_o(\omega_o)}\partial_{xx}A + i\frac{k_o(\omega_o)}{2}(n(x)^2 - 1)A$$

$$+ \sum_{m=2}^{3} i^{m+1}\frac{\beta_m}{m!}\partial_t^m A + A\frac{g_o(\omega)}{1 + \frac{\int |A(x,z,t')|^2\,dt'}{I_{gs}}}$$

$$+ i\gamma(1 + \frac{i}{\omega_o}\partial_t)[(1 - f_R)|A|^2 A + f_R \int_{-\infty}^{t} d\tau h_R(\tau)|A(x,z,t-\tau)|^2]$$

where $k_o(\omega_o)$ is the propagation constant at the center angular frequency $\omega_o$, n(x) is the index profile, $\beta_m$ are the m[th] order chromatic dispersion terms, $g_o(\omega)$ is the small-signal gain spectrum, $\gamma = n_2 k_o$ is the nonlinear coefficient, $f_R$ is the Raman fraction and $h_R(\tau)$ is the Raman response of the medium. The diffraction, dispersion and gain terms are numerically implemented in the frequency domain. To improve calculation speed, except where noted otherwise, $f_R = 0$ and the shock term proportional to $\frac{i}{\omega_o}\partial_t$ is neglected.

Full accounting of gain saturation (*i.e.*, not using a Taylor expansion for the 4[th] term above) is expected to be important for stable solutions and a full spatiotemporal saturable absorber, *i.e.*

$$A(x,z,t) \rightarrow A(x,z,t)\sqrt{1 - \frac{mod\_depth}{1 + \frac{|A(x,z,t)|^2}{I_{sat}}}}$$

is expected to have a potential role in stable spatiotemporal mode-locking. Here we are implicitly assuming an instantaneous saturable absorber. These effects cannot be easily incorporated into a set of coupled mode equations such as the GMMNLSE and furthermore, it is of interest whether pulses are stable in the presence of all guided modes, rather than merely a subset. Therefore, full-field spatiotemporal simulations are important to supplement the coupled-mode model described previously. However, given that many round trips constitute roughly $10^3$ m of propagation and the required longitudinal step size is roughly $10^{-5}$ m, to conduct simulations in a reasonable timeframe we have here chosen to simplify the physics by neglecting one of the spatial dimensions. A similar approach has proven to have qualitative and even surprisingly quantitative accuracy in studies of solitons in graded-index waveguides (*11, 31-33*). Overall, this model captures all the relevant physics and confirms that spatiotemporal mode-locking in the second cavity has many of the same features as in the first.

Fig. S17 shows the ideal index profile for the fiber used in experiments, while the one-dimensional profile was used for the simulations.

Fig. S18 shows different regimes of spatiotemporal mode-locking for increasing pump power. We observe the emergence multi-round trip spatiotemporal pulse evolutions, and then eventually a stable Gaussian output beam at high power. An animated version of this figure is included at the end of Supplementary Movie 1.

Fig. S19 shows the steady-state intracavity pulse evolution corresponding to the maximum gain saturation energy of the sequence shown in Fig. S18. These results show that even though the output beam is a stable Gaussian, the intracavity pulse evolution is spatiotemporal, and the field comprises multiple transverse mode families within the cavity. These results also show that STML in this cavity follows similar principles as in the first cavity, with spatial and spectral filtering helping to establish 3D periodic boundary conditions for the steady-state pulse. An animated version of this figure is included at the end of Supplementary Movie 1.

Fig. S20-21 show numerical predictions of high-power Gaussian pulse formation in a cavity using a fiber double the size of the one used in our experiments and prior simulations.

Figs. S22-25 show examples of more complex spatiotemporal pulse evolutions. These examples are included only to show a small sample of other phenomena that can occur in a multimode mode-locked laser cavity, and are not of major importance to the key conclusions of this manuscript. Nonetheless, we expect many readers will find them interesting. Figs. S22-23 show multi-round-trip and chaotic dynamics that occur when the spatial filter is reduced well below the fundamental mode size. Figs. S24-25 show noise-like multimode pulses, and multimode dark solitons.

Peak power assessment

Assessing the peak power of the generated pulses is important in order to rule out large contributions from a continuous-wave (CW) background (which can sometimes be indicated in the spectrum), and to confirm that the laser is not multi-pulsing or producing small sub-pulses. In combination with the many other measurements taken of the output field (see Methods, Fig. S1, e.g.), this helps to confirm the stable spatiotemporal mode-locking we observe. However, measuring the peak power of even standard single mode fiber oscillators is non-trivial. We use two approximate checks that confirm that the peak power is within a factor of 2, and so provide another good indication that CW backgrounds and multipulsing are not significant.

Approach 1: spectral broadening in single-mode fiber
A common approach is to launch sampled portions of the pulse train into single-mode fiber and compare the measured spectral broadening from self-phase modulation with numerical simulations. For the pulses in the multimode fiber lasers, we observe different spectra depending on how the light is coupled into the single-mode fiber, reflecting the spatiotemporally-complex nature of the pulse. However, because other measurements which average over the entirety or most of the field (spectrum, pulse train, interferometric autocorrelation, RF spectrum) imply that these variations are small over some pulses, we expect this approach is still valuable as a rough estimate or "sanity check." Here we apply this method to two representative examples of STML states where the output beam profile is relatively simple, and so we are relatively confident that the autocorrelation's implied pulse duration is a reasonable estimate for any given spatial sample of the pulse. Fig. S30-31 show two examples of this test being applied.

Approach 2: analytic estimate of intracavity nonlinear phase shift
In addition, since the spectral breathing ratio of the pulse evolution is known (the ratio of the output bandwidth to the bandwidth of the spectral filter), a second check can be made by comparing the *expected* peak nonlinear phase shift of the pulse, assuming formation of either dissipative solitons or similariton pulses, with the value calculated assuming the output pulse energy corresponds to a single, background-free pulse, and knowing the fiber length, the output mode area and the estimated gain and loss of the cavity.

For ~40-50 MHz dissipative soliton fiber cavities, the correlation of nonlinear phase shift with spectral breathing ratio is well-established (*41*), and meanwhile for similaritons the peak nonlinear phase shift accumulated in the gain fiber can be shown analytically to be:

$$\varphi_{NL} = \frac{3\gamma}{8}\left(\frac{gU_{in}}{\sqrt{\beta_2\gamma/2}}\right)^{2/3}(\frac{\omega_p}{\omega_o})^2$$

The additional nonlinear phase in other fibers of length L can be estimated as $\frac{L}{L_{nl}}$, where $L_{nl} = \frac{A_{eff}}{k_o n_2 P}$, where $P$ is the peak power of the chirped pulse, $A_{eff}$ is the effective mode area, $k_o$ is the vacuum wavenumber and $n_2$ is the nonlinear index. We assume weak spatial breathing (the output mode-field diameter is roughly the same through the cavity), output coupling in the typical range of 30-70%, and take into account the 9 nm filter bandwidth and typical measured chirped duration and spectral bandwidth of the output pulse.

As an example, we consider the mode-locked state at 30 W pump power in Figure 3. Similar analysis can be carried out for all the other mode-locked states we observe, which yield quite similar conclusions. The output spectrum combined with the 9-nm filter corresponds to a spectral breathing ratio of ~1.5-2. The spectrum and breathing are most consistent with dissipative soliton pulse shaping, so we expect by comparison with (*34*) that if the laser is single-pulsing, and the output field consists only of the mode-locked pulse, the nonlinear phase shift is 3-4π.

To assess, we assume the measured average power divided by the repetition rate of the measured pulse train gives the energy in the single pulse. From this, we estimate the nonlinear phase shift to confirm whether it is consistent with, or contradicts, the expected value above. A Gaussian fit of the output beam profile gives an effective mode area ($A_{eff}$) of 3850 µm² and the output pulse duration is ~1.5 ps. The average intracavity pulse energy is taken to be $E_{in,eff} = \frac{E_{out}}{OC} f$, where $f = \frac{1}{LE_{final}} \int_0^L E(z) dz$ and E is the pulse energy and $E_{final}$ is the energy before output coupling. Taking the cavity-averaged peak power to be $P = \alpha E_{in,eff}/T_{out}$, where $\alpha$ is a constant factor near 0.94 for a Gaussian pulse, the nonlinear phase shift is $\frac{L}{L_{nl}}$, where $L_{nl} = \frac{A_{eff}}{k_o n_2 P}$ and $L$ is the fiber length in the cavity. From this, we can write that $\varphi_{NL} = \frac{L k_o n_2 \alpha f E_{out}}{A_{eff} T_{out} OC}$. For the 3.6 m long fiber in the cavity, and taking $n_2 = 2.3 \cdot 10^{-20} m^2/W$, we obtain $\varphi_{NL} \approx 11.4 \alpha f / OC$.

Reasonable estimates for $f$ are in the range 0.3-0.8, which is the same as the expected values of $OC$ (however, if $OC$ is higher, $f$ is expected to be lower). Taking 0.3 and 0.8 as the min/max values for $OC$ and $f$, and $\alpha = 0.94$, the estimated nonlinear phase shift is $\varphi_{NL} \approx (5 \pm 2)\pi$, which is consistent with the expected 3-4π value. It is interesting that this shift exceeds the expected value for single-mode dissipative solitons, which would be expected for MM dissipative solitons. However, both methods described here provide confirmation of the peak power to within a factor of 2 or so, so the apparent quantitative agreement should not be taken as significant.

Discussion of spatiotemporal dispersion in spatiotemporally mode-locked pulses
In this section we examine numerically the spatiotemporal properties – the dispersion or space-time coupling – present in spatiotemporally mode-locked pulses. For the majority of situations we consider in this work, these effects are small. In some situations, they account for our inability to produce transform-limited autocorrelations of complex spatiotemporal pulses using just quadratic chromatic dispersion compensation. For the purposes of the current manuscript, this is the main conclusion of this section. However, the presence of spatiotemporal dispersion represents an important distinction from conventional mode-locked pulses, which may be important for future developments. Therefore, more detail is considered here.

For normal dispersion mode-locking in single-mode fiber, dissipative solitons and similaritons in general have a substantial chirp. For single-mode lasers, this can be removed by a diffraction grating compressor to produce nearly transform-limited pulses. For the multimode pulses formed here, we expect not only a chirp in time, but also a spatiotemporal pulse structure due to spatial, or modal, dispersion. Consideration of spatiotemporal pulse dispersion is not, however, novel. All dispersive, spatially inhomogeneous optical components (e.g., diffraction grating compressors, lenses, etc.) introduce significant spatiotemporal coupling, or spatiotemporal dispersion. For example, focusing by a lens – which is a spatially inhomogeneous dispersive element – leads to a focus in which a conventional mode-locked pulse is, without prior compensation, spread out in spacetime over 100-1000 fs. An excellent overview is (*35*) and its references. This has been the main driver for the development of spatiotemporal pulse measurement tools historically, and consideration of spatiotemporal coupling is necessary for use of ultrashort pulses in general.

Most spatiotemporally mode-locked states we observe, in simulation and experiment both, exhibit only small spatiotemporal dispersion, and in the context of conventional applications we expect this to be of no greater importance than for conventional mode-locked oscillators. Remarkably, this is even true when the intracavity field comprises ~100 spatial modes, such as in Fig. S14. This observation may be related to the condensation-like nonlinear attraction to low-order modes, 'beam cleanup', that has now been observed in many works studying multimode GRIN fibers *(12-16)*. This is confirmed by the results below, and experimentally by the nearly transform-limited autocorrelations and peak-power. However, this is not true of spatiotemporally mode-locked pulses in general, and spatiotemporal dispersion means that certain multimode mode-locked states cannot be compressed to their transform limit without some kind of spatiotemporal dispersion control outside the cavity.

Figs. S32 shows the spatiotemporal dispersion properties for a spatiotemporally mode-locked pulse (also shown in Figs. S5-8) which is similar in spatial mode content to typically-observed spatiotemporally-mode-locked pulses. This examples show that, for many spatiotemporally mode-locked pulses, full quadratic spatiotemporal dispersion compensation can improve the pulse quality, but has an overall small effect. This explains why we can observe nearly transform-limited autocorrelations experimentally without deliberately controlling spatiotemporal dispersion.

Figs. S33 considers a spatiotemporally mode-locked pulse for which conventional dispersion compensation and simple spatiotemporal dispersion compensation are insufficient to reach the transform limit (these very multimode pulses are also shown in Fig. S9-12). These results are similar to some observed experimentally.

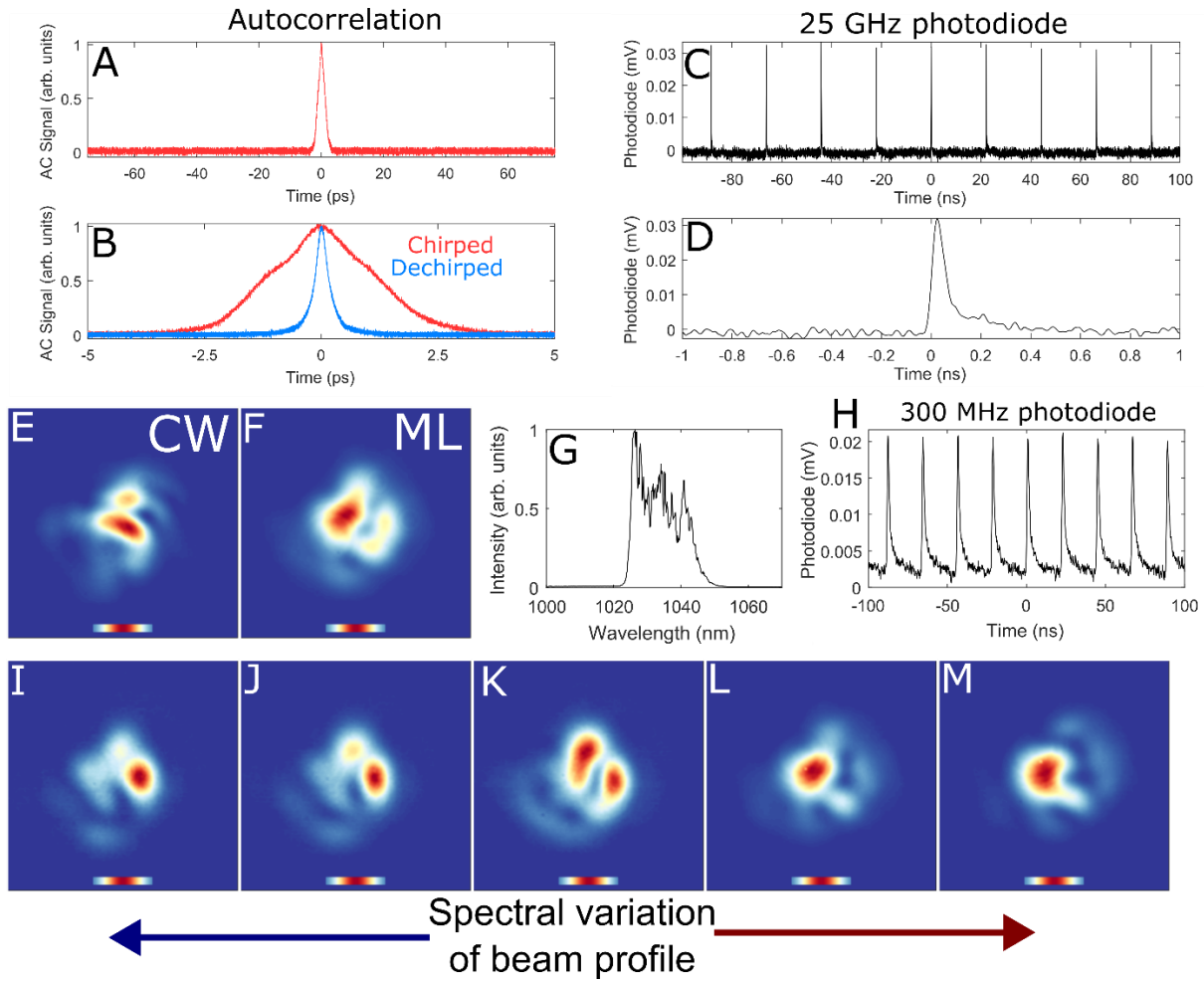

**Fig. S1: Measurements verifying spatiotemporal mode-locking.**

After mode-locking the laser, we measured the long-range intensity autocorrelation of the chirped and dechirped pulses (**A,B**). The autocorrelation trace shows no multiple or sub-pulses over the entire >185 ps range. Together with fast photodiode measurements with ~ 40 ps resolution (**C,D**), we conclude the laser is producing only one pulse per round trip. For many measurements, a slower photodiode must be used (**H**). Due to the slower decay of this photodiode, a DC-offset appears. However, based on **A-D**, we conclude that there is virtually no continuous wave or amplified spontaneous emission background. **G** shows the mode-locked laser spectrum. **E** and **F** show the near-field beam profile of the laser operating in continuous wave (CW) mode slightly below the mode-locking (ML) threshold, and slightly above. To further verify the multimode nature of the mode-locked pulse, we measure the beam profile at different regions of the spectrum by rotating a narrowband filter (**I-M**). The profiles are measured at roughly equally spaced intervals across the 20 nm bandwidth. These show that all spectral regions of the mode-locked pulse is multimode, but with varying modal phases/energies. Furthermore, no regions of the spectrum show a beam profile that resembles the CW operation.

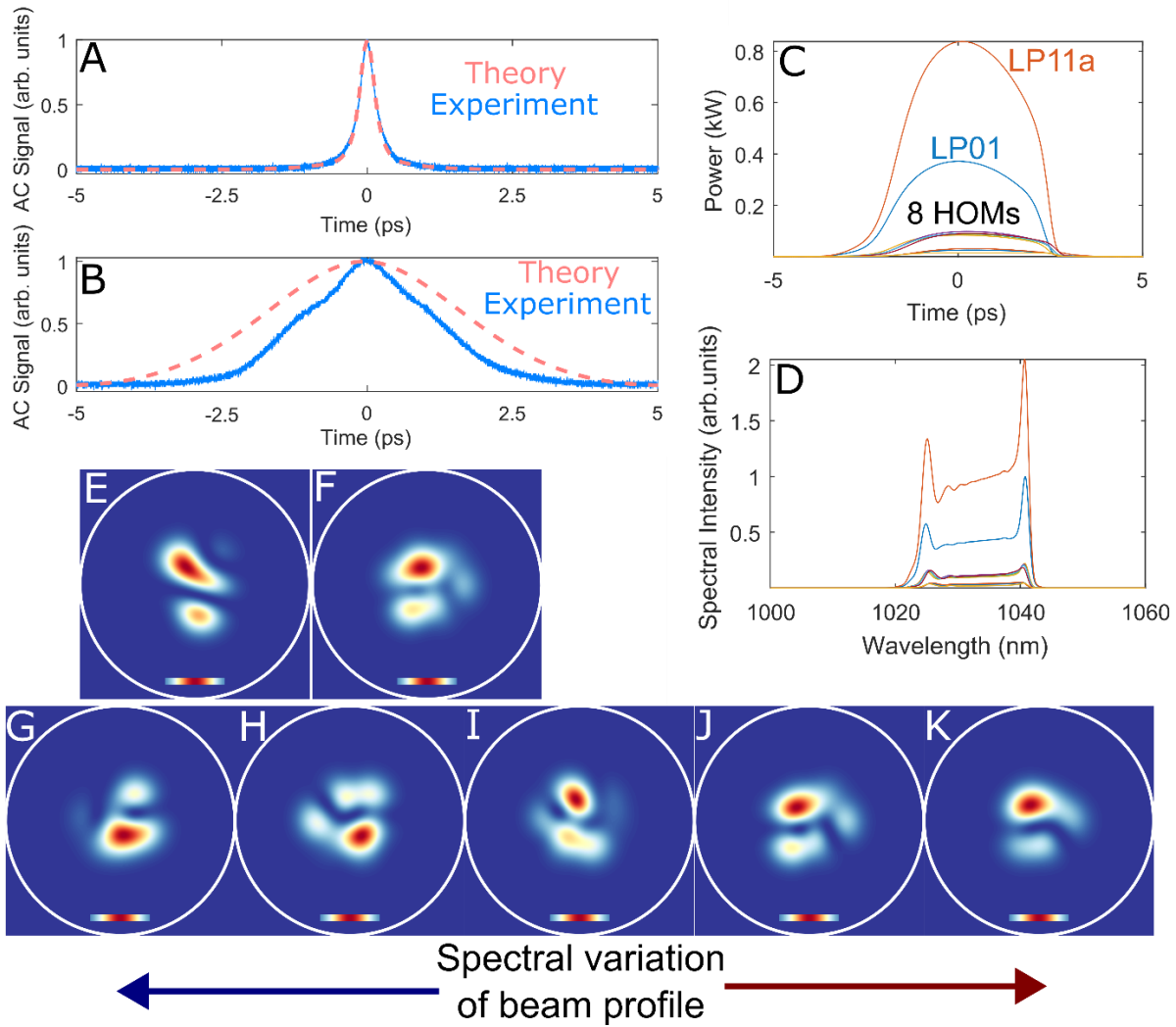

**Fig. S2: Theoretical modelling of STML shown in previous figure.**
**A** (**B**) shows the dechirped (chirped) intensity autocorrelation for both theory and experiment. Experimentally, the shape of this kind of autocorrelation varies with alignment (in simulations, we observe that the temporal shape of the pulse varies in space). Therefore, for **A** and **B** we chose to compare the alignment/position for which maximum signal is expected. **C** shows the mode-resolved output pulse intensity, while **D** shows the mode-resolved spectral intensity. **E** shows the predicted CW lasing beam profile, which is a narrowband electric field with modal coefficients given below. **F** shows the simulated steady-state mode-locked beam profile (averaged over all frequencies), while **G-K** show samples of the beam profile at frequencies roughly equally spaced across the spectrum. Since many parameters in the simulations cannot be directly measured, the results above represent our best estimates. Although we expect that significantly better agreement could be obtained by fine-tuning these parameters, we note that the qualitative agreement is excellent, and the quantitative agreement is also good.
**Modelling details:** The model used is described in the manuscript and Supplementary Section 2. Since we cannot directly measure the coupling coefficients between the gain fiber and passive fiber, and the coupling of each mode into the gain fiber, to model the experiment, we estimated these coefficients to be in proportion [4 6 2 2 2 2 2 1 1 1] (the total magnitude is then normalized to 1), where the coefficients refer to each of the 10 lowest-order modes of the passive GRIN fiber. That is, in coupling from the gain fiber to the passive fiber, we expect excitation of a field with an electric field amplitude in mode 1 (which is LP01) that is initially twice that of LP11b and the next 4 higher order modes, and amplitude in mode 2 (which is LP11a) that is initially three times that of LP11b and the next 4 higher order modes. We expect coupling from free-space into the gain fiber to be described by the same coefficients. The similarity of these values is not a necessity, as in general the overlap between the field at the gain fiber, and overlap of the gain fiber with the GRIN fiber, can be different. However, they are taken to be the same here since this corresponds to the highest-efficiency CW lasing operation, which corresponds to the cavity configuration we examined experimentally in the previous figure. The rest of the cavity parameters were chosen to be the experimentally-measured values, except for the modulation depth and saturation power of the saturable absorber, the output coupling ratio and net additional loss, and the gain fiber's small-signal gain coefficient and the gain saturation energy. These latter values were estimated to be 100%, 2 kW, 70%, 50%, 30 dB, and 2 nJ, since these are typical parameters used in accurately modelling normal-dispersion mode-locked fiber lasers based on single-mode fiber. 100% modulation depth is not a requirement for mode-locking; it is used here in order to minimize the starting time from noise, given how computationally intensive the simulations are. While more investigation is needed, it is likely that a high modulation depth saturable absorber (>50%) will usually be required, similar to normal dispersion mode-locked oscillators in single-mode fiber.

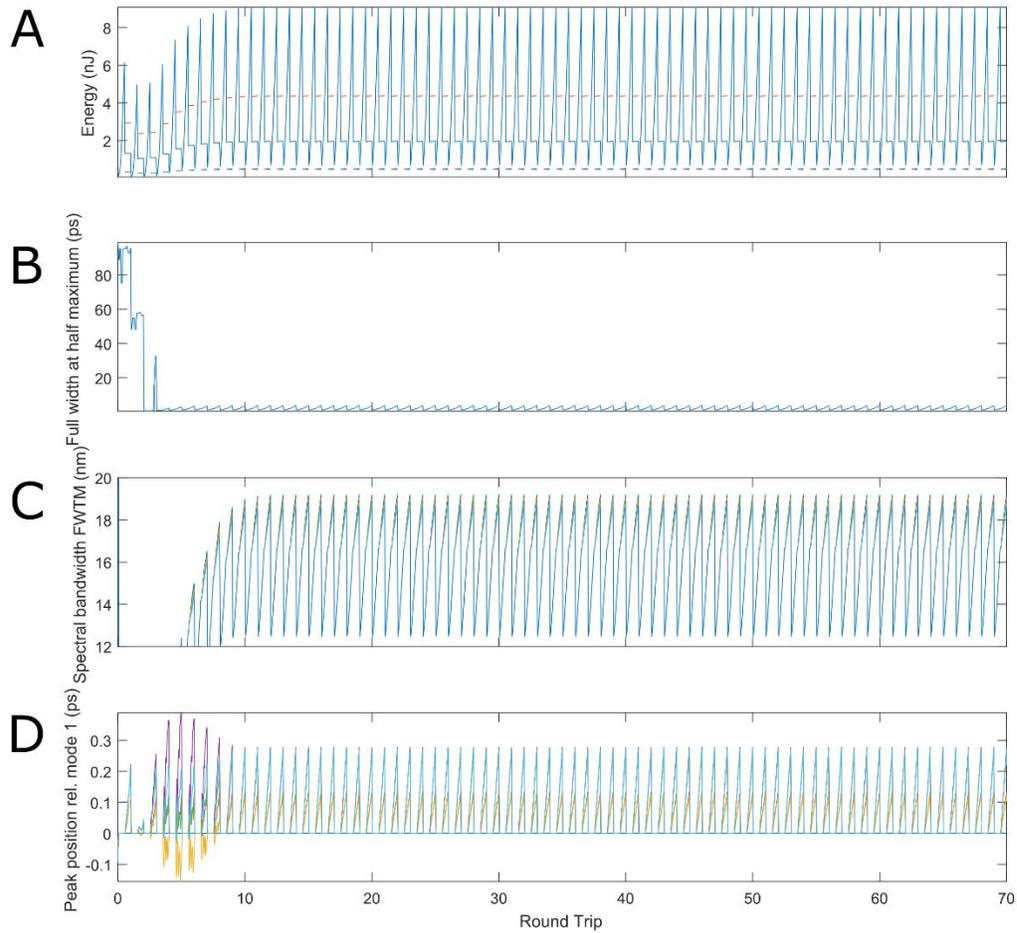

**Fig. S3: Intracavity, mode-resolved pulse evolution starting from a noise field.**
A shows the energy in each mode (colors are the same as in Fig. S2), B shows the pulse width around the most-intense pulse in each mode, C shows the spectral bandwidth (full width at 10% of maximum), and D shows the temporal peak position in each mode. After a short transient, we see that a steady-state periodic evolution is established within the cavity, with the pulse oscillating in space, time, spectrum and total energy over the course of each round trip.

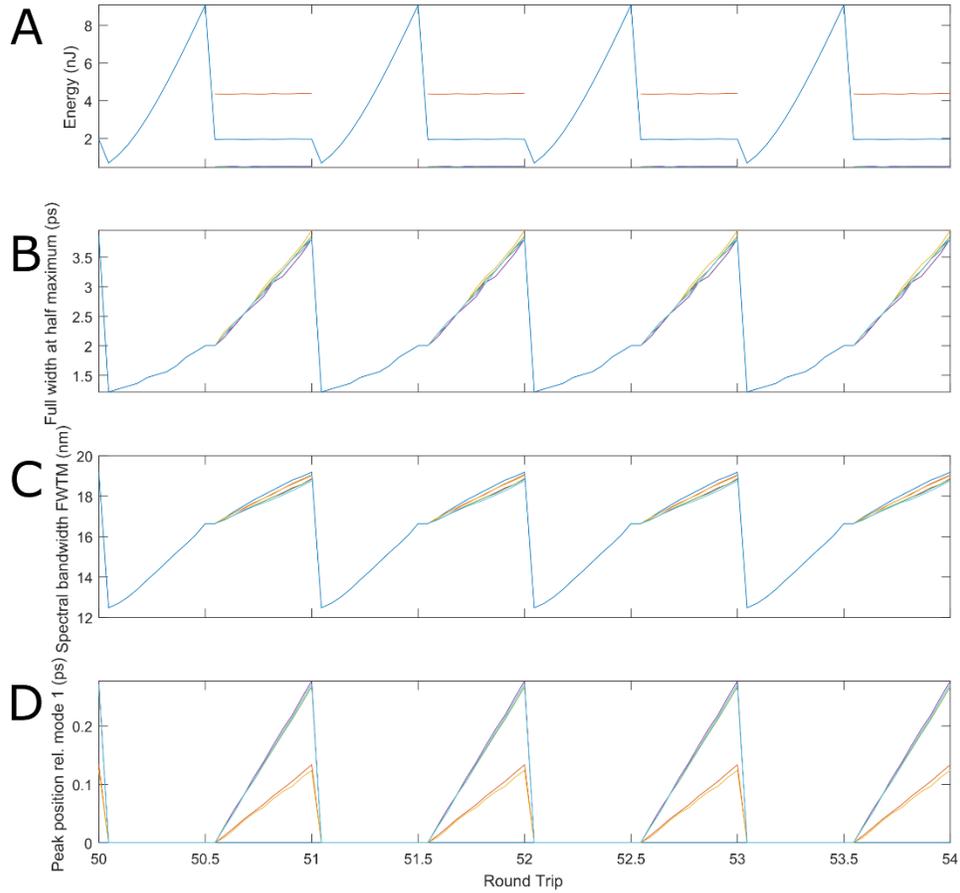

**Fig. S4: Intracavity, mode-resolved pulse evolution starting from a noise field, zoomed to show steady-state pulse evolution.**
**A** shows the energy in each mode (colors are the same as in Fig. S2), **B** shows the pulse width of the pulse in each mode, **C** shows the spectral bandwidth (full width at 10% of maximum), and **D** shows the temporal peak position in each mode. In **D**, note that because the modes fall into different groups, several curves nearly overlap. The round trip features first output coupling, saturable absorption, spectral filtering and coupling into the gain fiber (modelled as single-mode, see later sections), which provides the spatial filtering, followed by propagation through the gain fiber (0 to 0.5 round trips above), and then by propagation in the passive GRIN fiber (0.5 to 1 round trips above). The figure show that the pulses exhibit monotonic growth in pulse energy, bandwidth and pulse duration. To establish a self-consistent periodic evolution, these are cut by the output coupling, the spectral filter, and the spatial and spectral filter. Roughly speaking, over each round trip, the spatial filter compensates for modal dispersion, while the spectral filter compensates for chromatic dispersion. Hence, the mode-locking can be viewed as a spatiotemporal, or multimode, generalization of single-mode normal dispersion mode-locking principles.

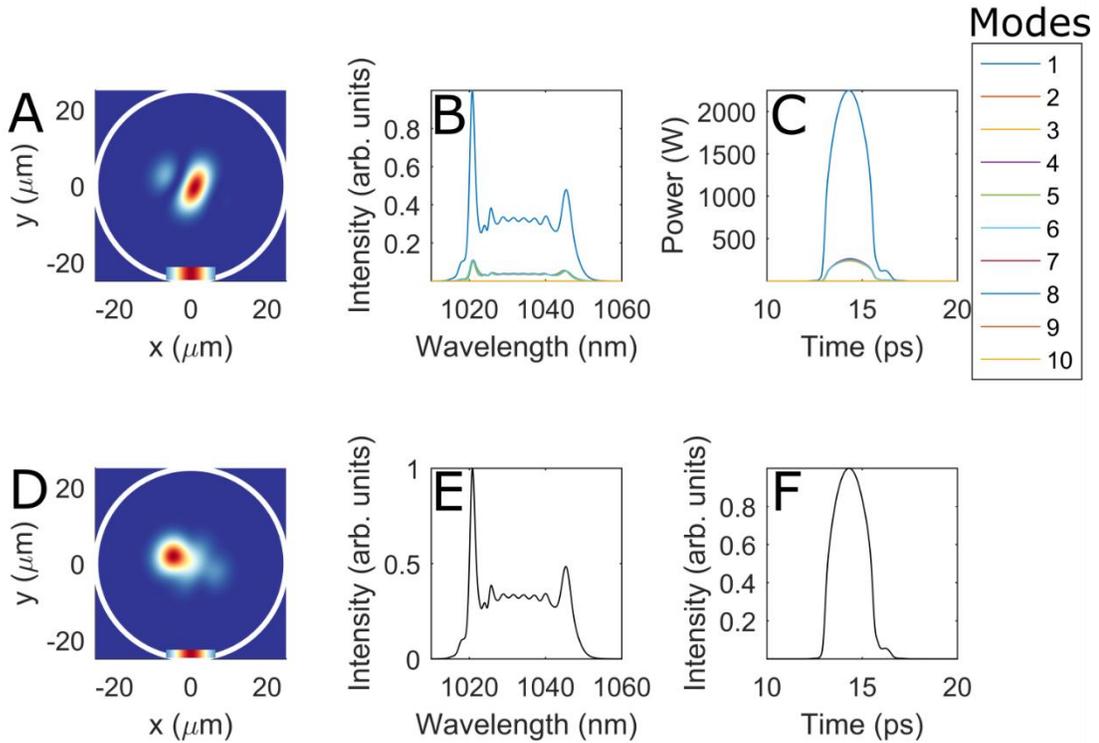

**Fig. S5: Typical numerical example of fundamental-mode dominant multimode mode-locking in the first cavity.**
From top left to bottom right: **A**. Expected continuous wave output profile, **B**. mode-resolved steady state spectra, and **C**. mode-resolved pulse intensity, **D**. steady-state output spatial profile, **E**. total steady-state spectrum and **F**. pulse intensity. The scale bars show the Gaussian intensity distribution of the fundamental mode, and the white circle indicates the fiber core. The legend in the upper right corner refers to the mode-resolved plots **B-C**. The 10 simulated modes are the first 6 lowest order modes, while modes 7-10 correspond to higher-order radially-symmetric modes which have a large overlap with the fundamental mode. The output coupling ratio is 80%, the modulation depth is 99% (with a saturation power 4 kW), the filter bandwidth is 9 nm, the gain saturation energy is 3 nJ, and the passive (gain) fiber length is 50 (150) cm. The projection coefficients for passive fiber to gain fiber and from gain to passive fiber are both set to [3 1 1 1 1 1 0 0 0 0] (the coefficients are subsequently normalized so that their total magnitude is 1 to conserve energy at each fiber transition). The continuous wave beam profile is estimated as the lowest-loss configuration of spatial modes, which corresponds to a field comprising exactly the projection coefficients just mentioned. As with the first example in Fig. S2-4, here and subsequently the choice of the same projection coefficients is not required, but is chosen since it corresponds to the experimental practice of aligning the cavity for maximal CW efficiency.

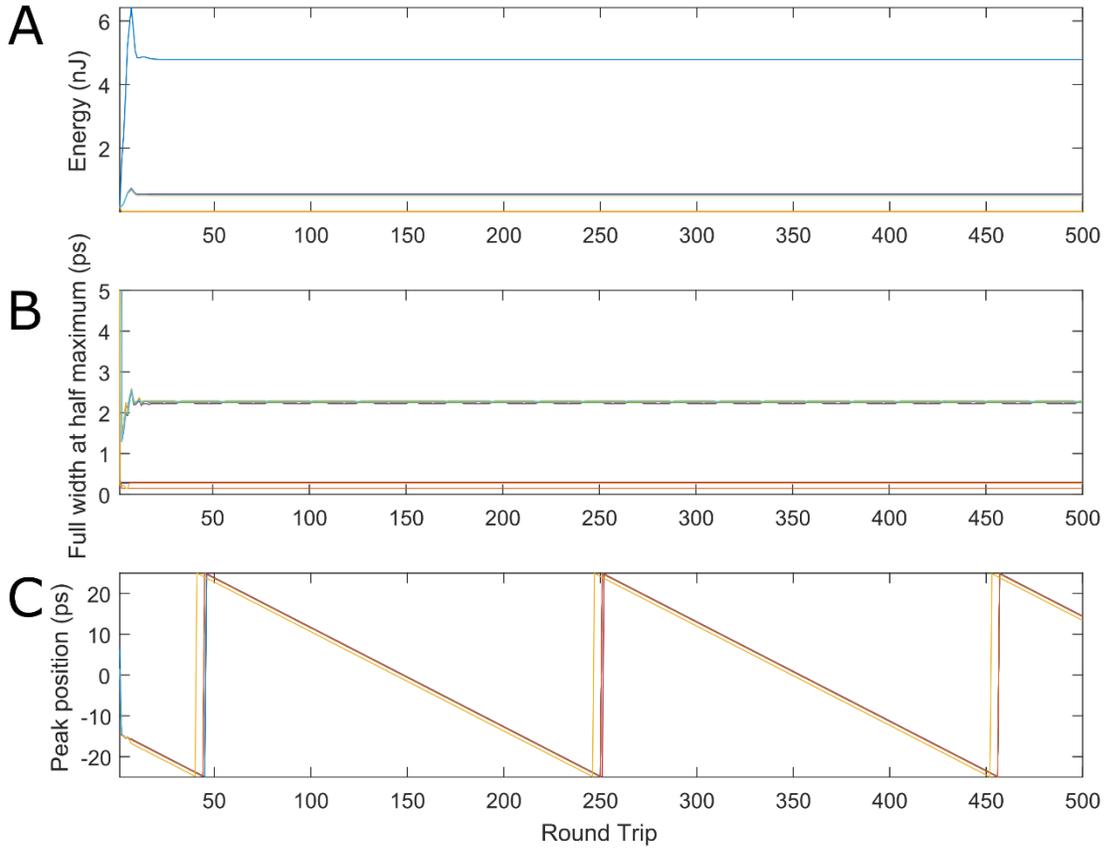

**Fig. S6: Output pulse parameters during formation and evolution of a multimode mode-locked pulse in the first (numerical results).**
Evolution of the output field in the low-energy cavity, corresponding to Fig. S5. The plots show the **A**. output pulse energy, **B**. duration, and **C**. the position of the pulse in time. After arising quickly from noise, the mode-locked pulses' properties are stable over many round trips and do not oscillate from round trip to round trip. Due in part to its spatially-multimode nature, the mode-locked pulse moves at a finite group velocity relative to the simulation window, which moves at the velocity of the fundamental mode. The artifacts at 50, 250 and 450 round trips are due to the pulse moving from one side (-25 ps) to the other side of the simulation window (25 ps).

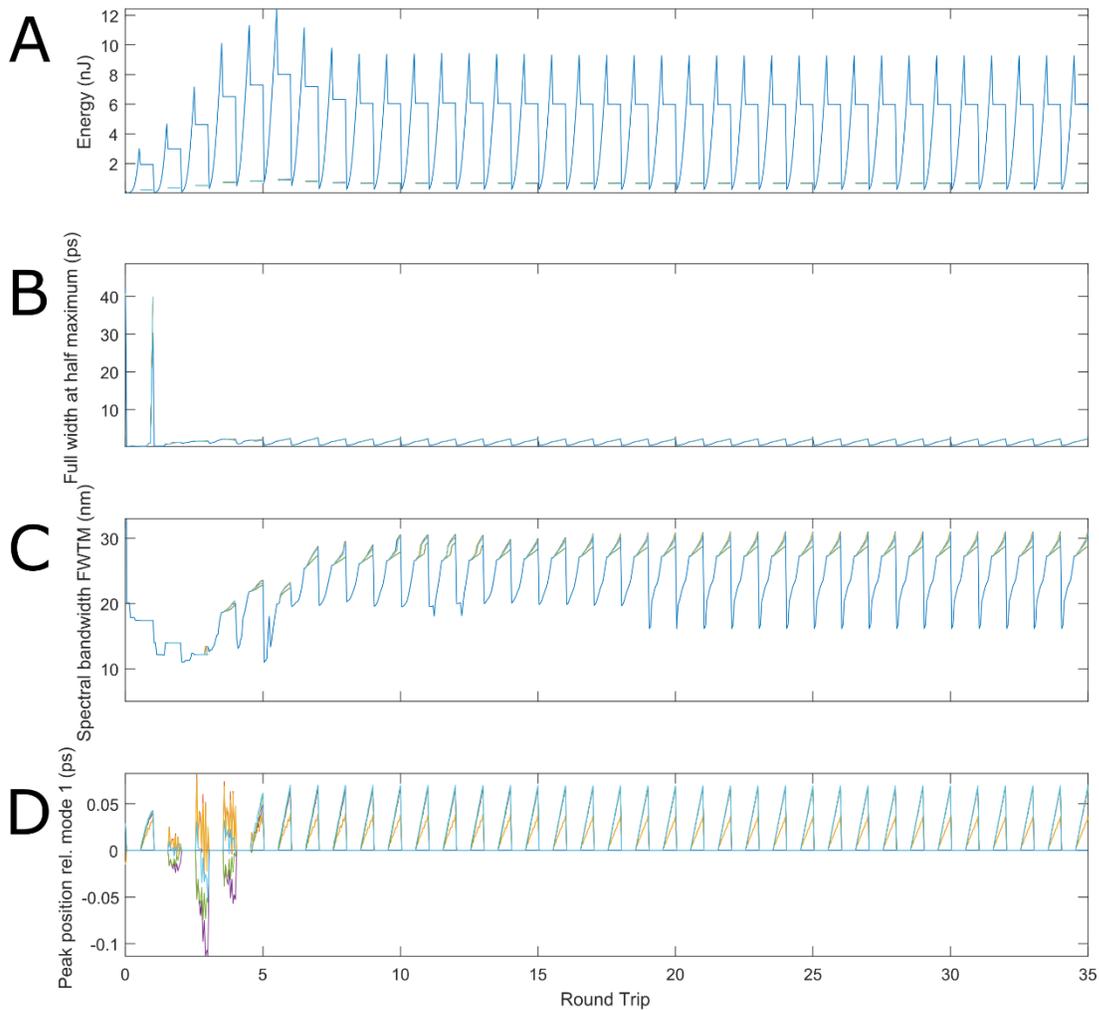

**Fig. S7: Starting intracavity pulse evolution corresponding to the previous two figures (numerical results).**
The figures show the evolution of **A**. the intracavity pulse energy, **B**. intracavity pulse duration, **C**. the spectral bandwidth (full width at 10% maximum) and **D**. the position of the pulse in each mode relative to the position of the fundamental mode pulse. The spatial and spectral filter act to restore the pulse energy, duration, bandwidth, and to undo modal dispersion, during each round trip. Different traces correspond to different modes, for legend see Fig. S5. Note that, since the modes fall into degenerate mode groups, several traces are nearly overlapped in the curves shown in **D**.

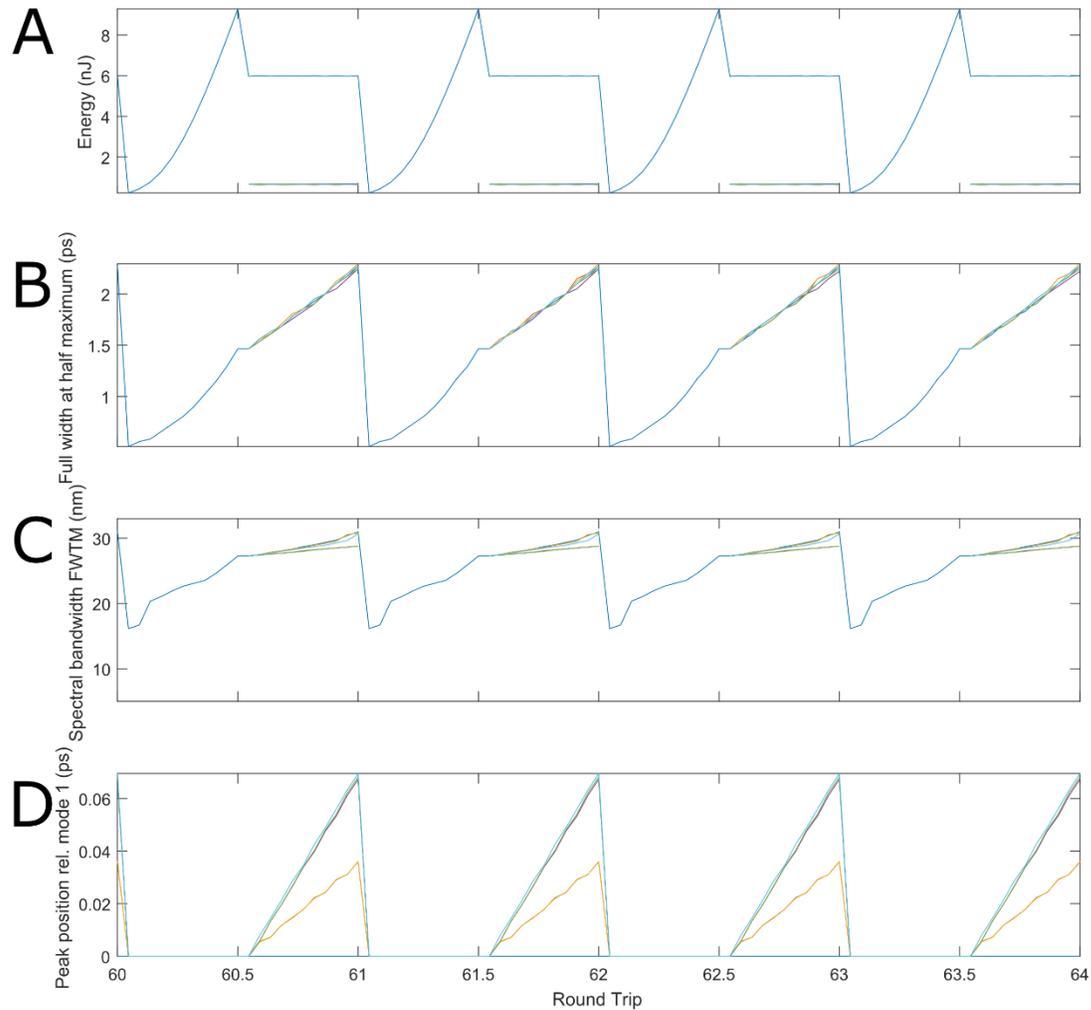

**Fig. S8: steady-state intracavity pulse evolution corresponding to the previous three figures (numerical results).**
The figures show the steady-state intracavity evolution of **A**. the pulse energy, **B**. pulse duration, **C**. the spectral bandwidth (full width at 10% maximum) and **D**. the position of the pulse in each mode relative to the position of the fundamental mode pulse. The spatial and spectral filter act to restore the pulse energy, duration, bandwidth, and to undo modal dispersion, during each round trip. Different traces correspond to different modes, for legend see Fig. S5. Note that, since the modes fall into degenerate mode groups, several traces are nearly overlapped in the curves shown in **D**. The figure shows mode-locking that is qualitatively similar to that described in Fig. S3-4.

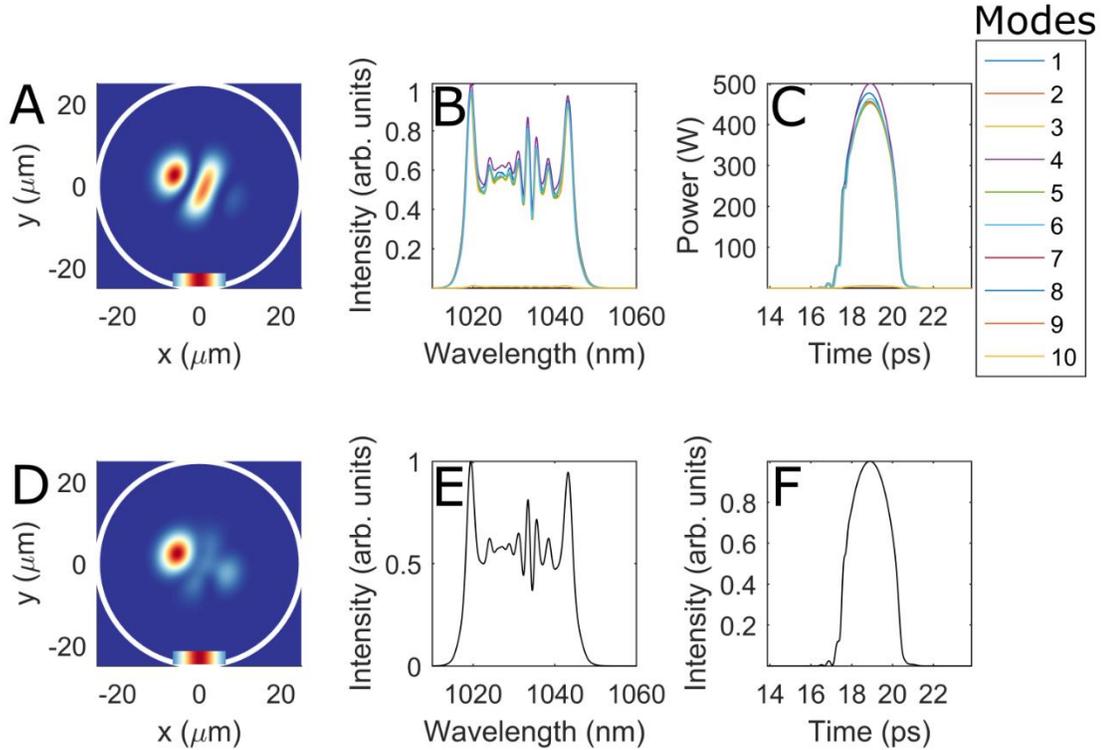

**Fig. S9: Typical numerical result for the first multimode cavity for multimode mode-locking with similar energy in many modes.**
From top left to bottom right: **A**. Expected continuous wave spatial output profile, **B**. mode-resolved steady state spectra, and **C**. mode-resolved pulse intensity, **D**. steady-state output spatial profile, **E**. total steady-state spectrum and **F**. pulse intensity. The scale bars show the Gaussian intensity distribution of the fundamental mode, and the white circle indicates the fiber core. The legend in the upper right corner refers to the mode-resolved plots **B-C**. Here, the simulated modes are the first 6 lowest order modes, while modes 7-10 correspond to higher-order radially-symmetric modes which have a large overlap with the fundamental mode. The output coupling ratio is 80%, the modulation depth is 99% (with a saturation power 4 kW), the filter bandwidth is 9 nm, the saturation energy is 3 nJ, and the passive (gain) fiber length is 50 (150) cm. The projection coefficients for passive fiber to gain fiber and from gain to passive fiber are both set to [1 1 1 1 1 1 0 0 0 0] (the coefficients are subsequently normalized so that their total magnitude is 1 to conserve energy at each fiber transition). The pulse is generally stable, but exhibits aperiodic collapse and revival.

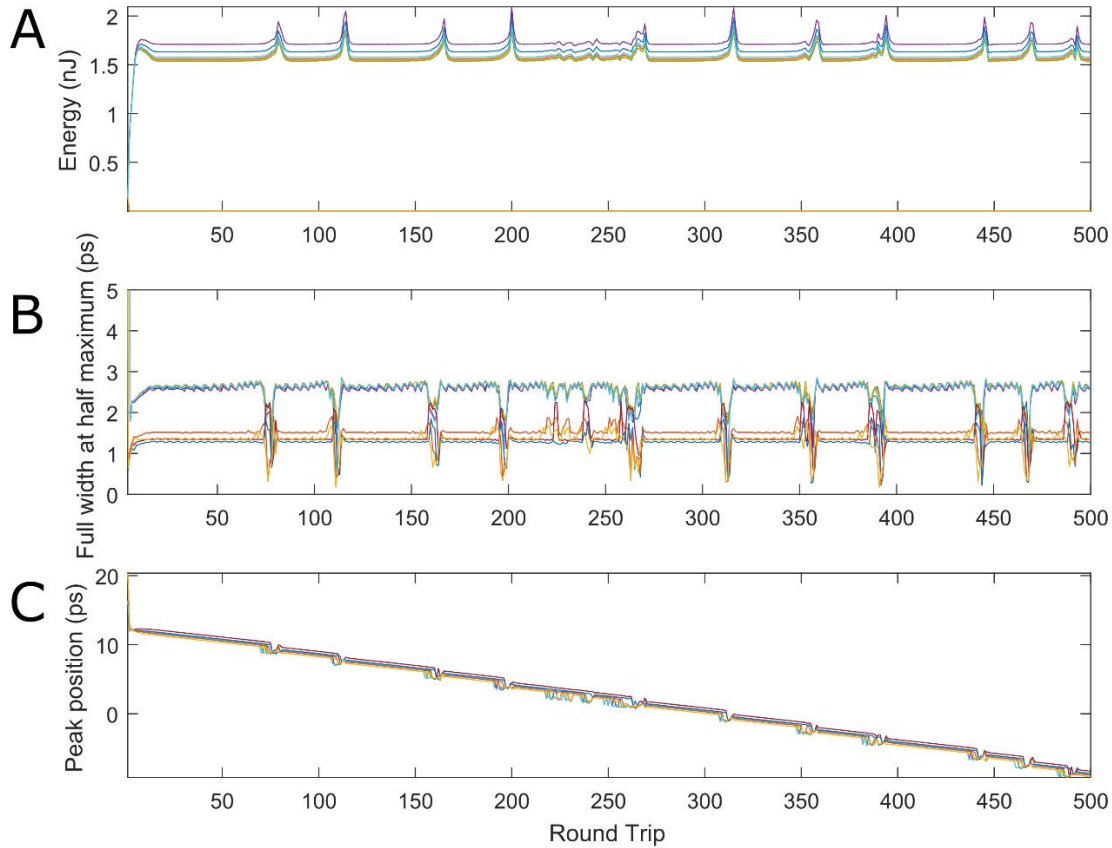

**Fig. S10: Evolution of the field for multimode mode-locking in the first cavity with similar energy in many spatial modes.**
Numerical results corresponding to Fig. S9. The plots show the **A**. output pulse energy, **B**. duration, and **C**. the position of the pulse in time. The mode-locked pulses' properties are stable over many round trips and do not oscillate from round trip to round trip. However, with no particular period, the pulse exhibits an instability and collapses, then quickly returns. Due in part to its spatially-multimode nature, the mode-locked pulse moves at a finite group velocity relative to the simulation window, which moves at the velocity of the fundamental mode.

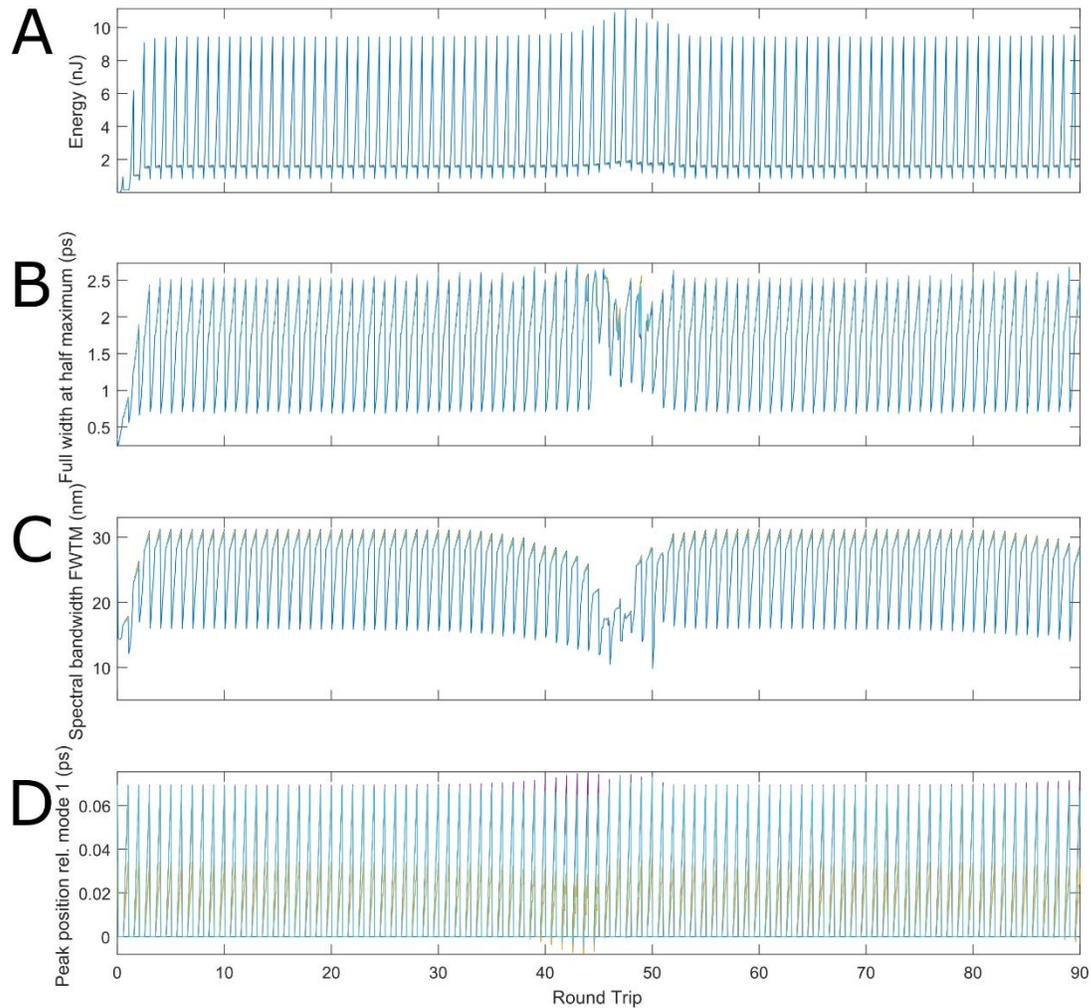

**Fig. S11: Intracavity, mode-resolved pulse evolution starting from a noise field.**
**A.** shows the energy in each mode (colors are the same as in Fig. S2, S5, S9), **B.** shows the pulse width around the most-intense pulse in each mode, **C.** shows the spectral bandwidth (full width at 10% of maximum), and **D.** shows the temporal peak position in each mode. After a short transient, we see that a steady-state periodic evolution is established. A short collapse/revival can be seen around 40-50 round trips. The figure shows mode-locking that is qualitatively similar to that described in Fig. S3-4.

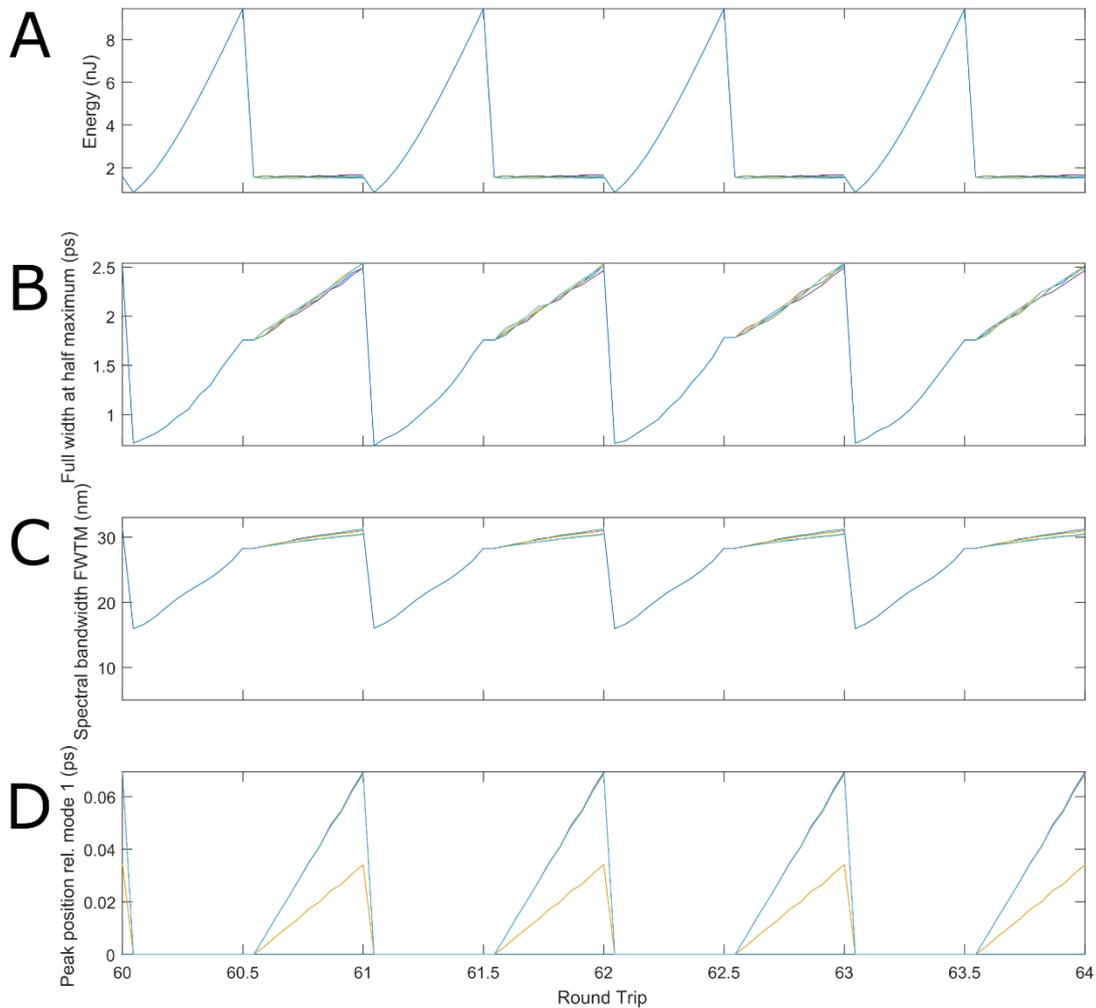

**Fig. S12: Intracavity, mode-resolved pulse evolution starting from a noise field, zoomed to show steady-state pulse evolution.**
**A.** shows the energy in each mode (colors are the same as in Fig. S2), **B.** shows the pulse width around the most-intense pulse in each mode, **C.** shows the spectral bandwidth (full width at 10% of maximum), and **D.** shows the temporal peak position in each mode. The figure shows mode-locking that is qualitatively similar to that described in Fig. S3-4.

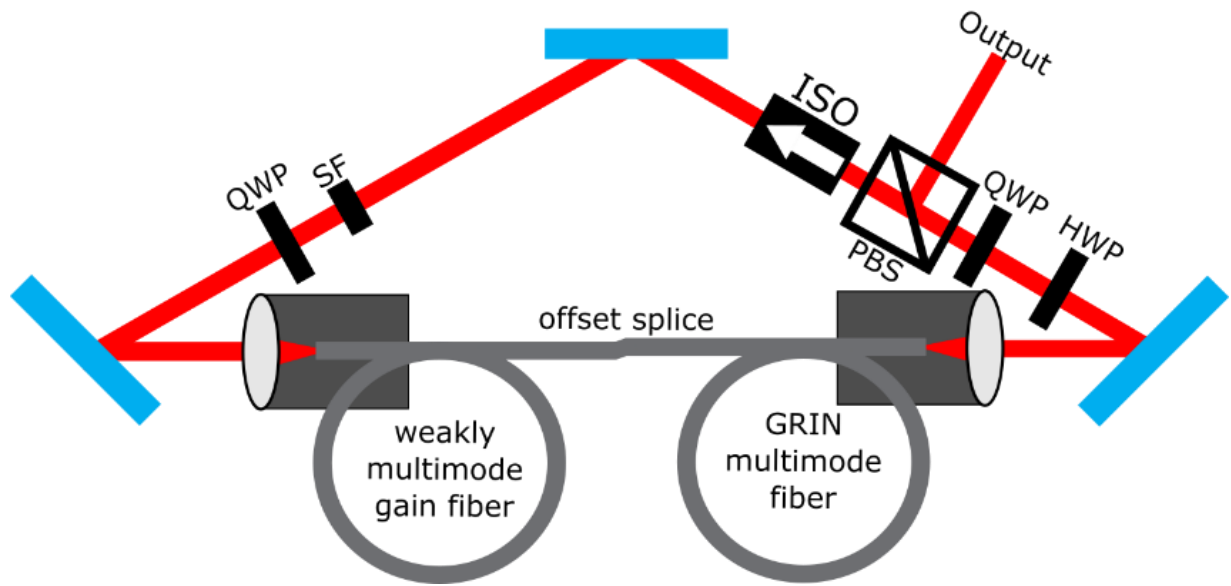

**Fig. S13: Schematic diagram of the cavity supporting spatiotemporal mode-locking with primarily ultrafast Kerr interactions.**
Highly multimode excitation of the graded-index passive fiber is accomplished by a spatially-offset splice from the few-mode gain to highly-multimode graded-index passive fiber. Spatial filtering is accomplished by the overlap of the field coupled into the gain fiber. Nonlinear polarization rotation is used as an effective saturable absorber, controlled by the output coupling polarizing beamsplitter (PBS) and the half-wave and quarter-wave plates (HWP and QWP). The isolator ensures unidirectional light propagation in the ring cavity. A single passband bandpas spectral filter (SF) is implemented using an interference filter.

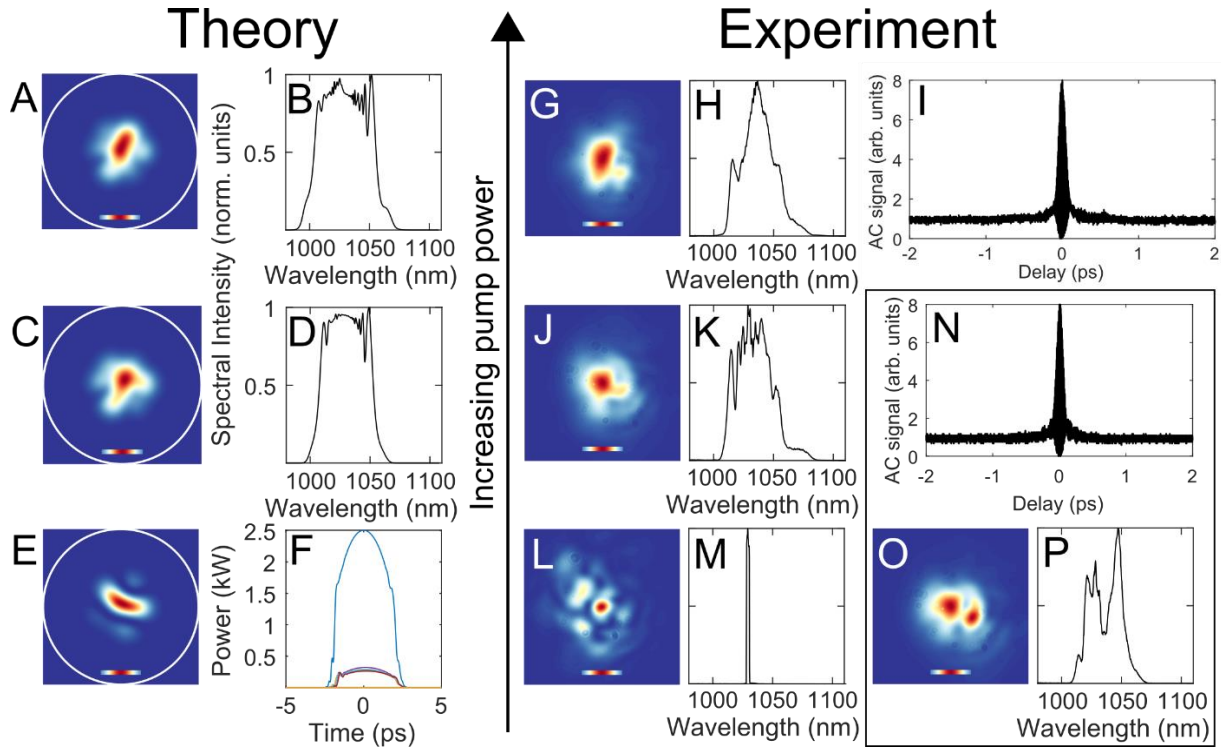

**Fig. S14: Highly multimode spatiotemporal mode-locking in the first cavity.**
These results illustrate mode-locking with a large number of transverse modes.

    **Numerical results A-F** provide qualitative modelling with a notably smaller set of modes. **A, C**. steady-state beam profile for simulation of a typical cavity for different gain saturation energies, **B, D**. corresponding total spectrum, averaged over 75 round trips. **E**. typical speckled beam profile expected for continuous wave lasing in the 10 transverse modes considered in the simulations. **F**. mode-resolved temporal intensity profile corresponding to **C-D** (different traces are different spatial modes). The pulse's ~14-nJ energy is comprised of ~8.5 nJ in the fundamental mode, and 1 nJ in the next 6 higher-order modes.

    **Experimental results G-P. G-I** show the beam profile, spectrum (linear scale) and autocorrelation of the dechirped pulse for a given cavity configuration. **J-M** show the beam profile and spectrum for the output field for decreasing pump power for the same cavity configuration. This sequence, along with **A-F**, illustrates the spatiotemporal transition from a highly-multimode continuous wave lasing to spatiotemporal mode-lockings with increasing pump power, and the spatiotemporal changes that occur as pump power is further increased beyond the mode-locking threshold. The simulation and experimental results are not directly comparable, since the simulation considers only 10 of the >100 transverse modes in the experiment, and because the experimental fiber-fiber mode coupling coefficients are not precisely known. When the cavity alignment is adjusted, different multimode pulses can be observed (**N-P**). The scale bar on all beam profiles shows the Gaussian intensity of the fundamental mode (13 µm mode-field diameter).

    **Analysis of Agreement between Simulations and Experiments.** As indicated above, the simulations were performed with 10 modes, which is clearly far fewer than the number of transverse modes relevant in the experimental results. The much higher transverse mode content of the experimental results is most clearly evident in comparing **E** and **L**. Since the only change from **L** to **J**, and **O** is a small increase in pump power, only ultrafast nonlinear interactions can

account for the change in the beam profile here, and therefore we conclude that the relatively-localized output beam must be a result of 'beam cleanup' occurring each roundtrip in the cavity. Since the saturable absorber rejection port is taken as the output, we conclude that Kerr-driven beam cleanup *(12-16)* is more important than saturable-absorber-related effects.

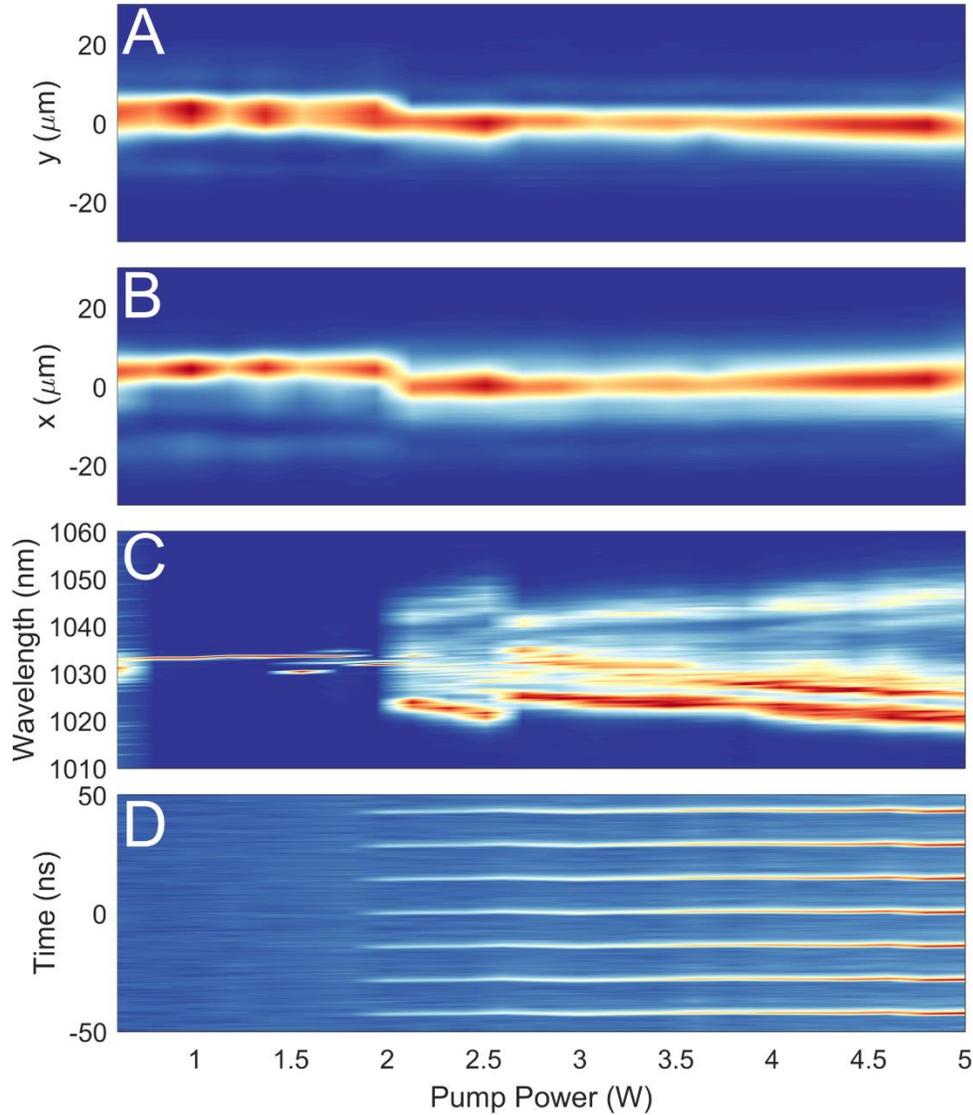

**Fig. S15: Measurement of spatiotemporal transition to mode-locking and spatiotemporal pulse shape evolution with pump power, in the first cavity.**

The plots show **A-B**. cross-sections of the near-field beam profile (x, y), **C**. the spectrum, and **D**. the pulse train. The transition to mode-locking is similar to the second cavity, shown in Fig. 3, but we are able to access interesting regimes and much stronger changes in the STML pulse properties above the mode-locking threshold. As the pump power is changed above the threshold, we observe changes in the beam profile and spectrum. As the pulse energy grows, the spectrum goes through a period of broadening (2-2.5 W), then suddenly shrinks. At this point the beam profile exhibits some subtle changes indicating a change in the modal composition of the pulse. Higher pump powers cause slow spectral broadening and spatial changes in the beam. Given the spatiotemporal nature of the pulse – its modal composition varies with wavelength and time – estimation of the mode composition is difficult. However, a reasonable guess is that the fundamental mode energy grows monotonically from ~2.5 to 5 W.

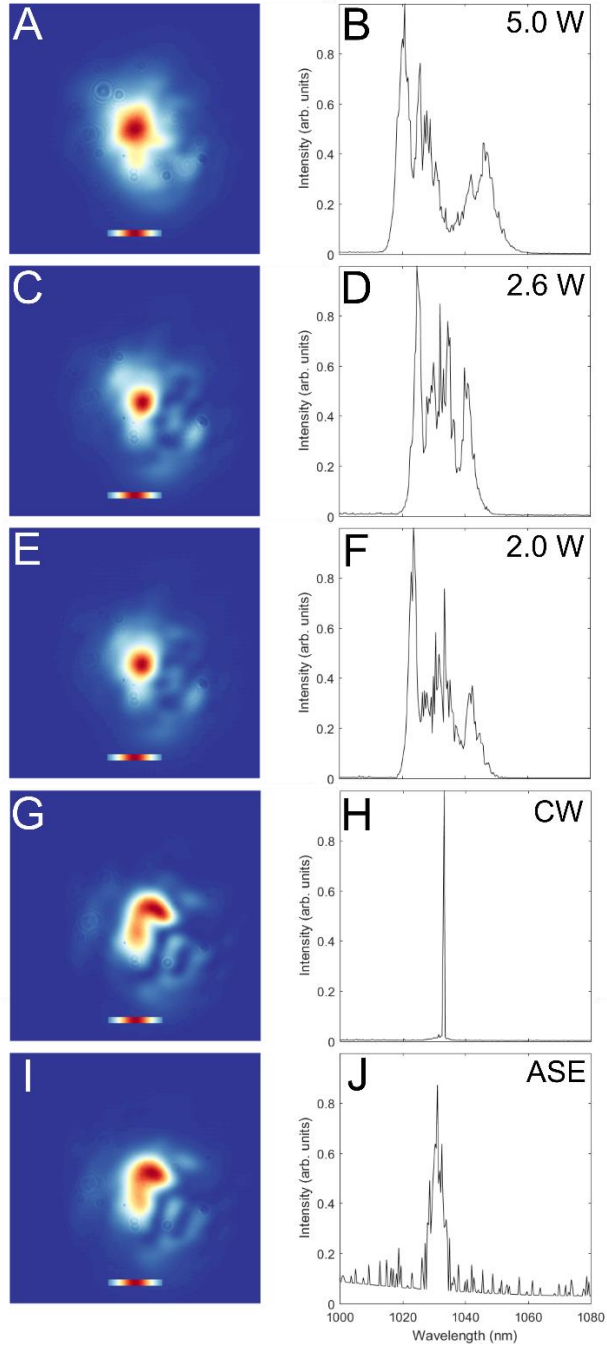

**Fig. S16: Near-field beam profiles and spectra corresponding to the indicated points in the previous figure.**
The plots show the measured beam profile and spectrum for the ASE (**I**, **J**), the CW lasing state (**G**, **H**), the first mode-locked state at 2 W pump power (**E**, **F**), the changed mode-locked state at 2.6 W (**C**, **D**), and the highest-power state shown at 5 W (**A**, **B**). The scale bar is the Gaussian intensity of the 13 µm mode field diameter fundamental mode.

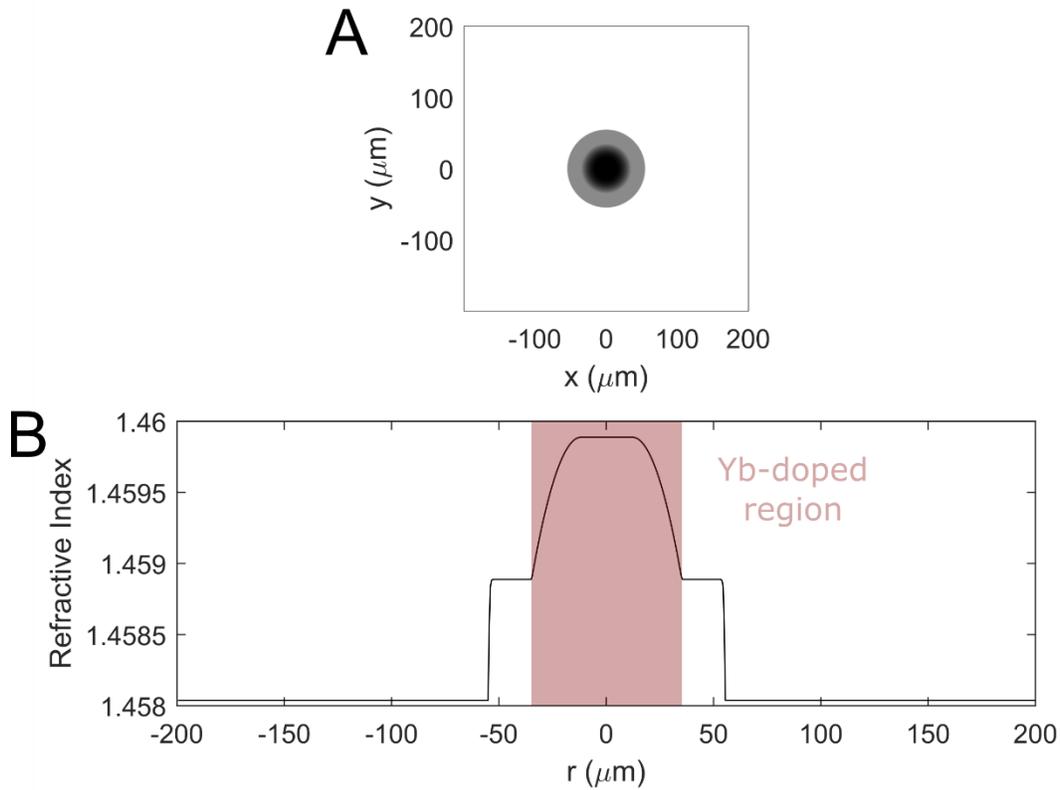

**Fig. S17: Index profile of the partially-graded multimode Yb-doped fiber used for the all-multimode fiber cavity.**
**A.** The 2D, radially-symmetric nominal profile of the fiber used in the experiments, **B.** a cross-section of **A.**, and the 1D index profile used in the (2+1)D simulations. For simulations, we assume for simplicity that the entire fiber is doped uniformly. However, in experiments the fiber is nominally doped only over the center 70 µm diameter region, shown in red. We expect this to have a quantitative influence, by contributing some additional effective spatial filtering, and affecting which modes receive the most gain.

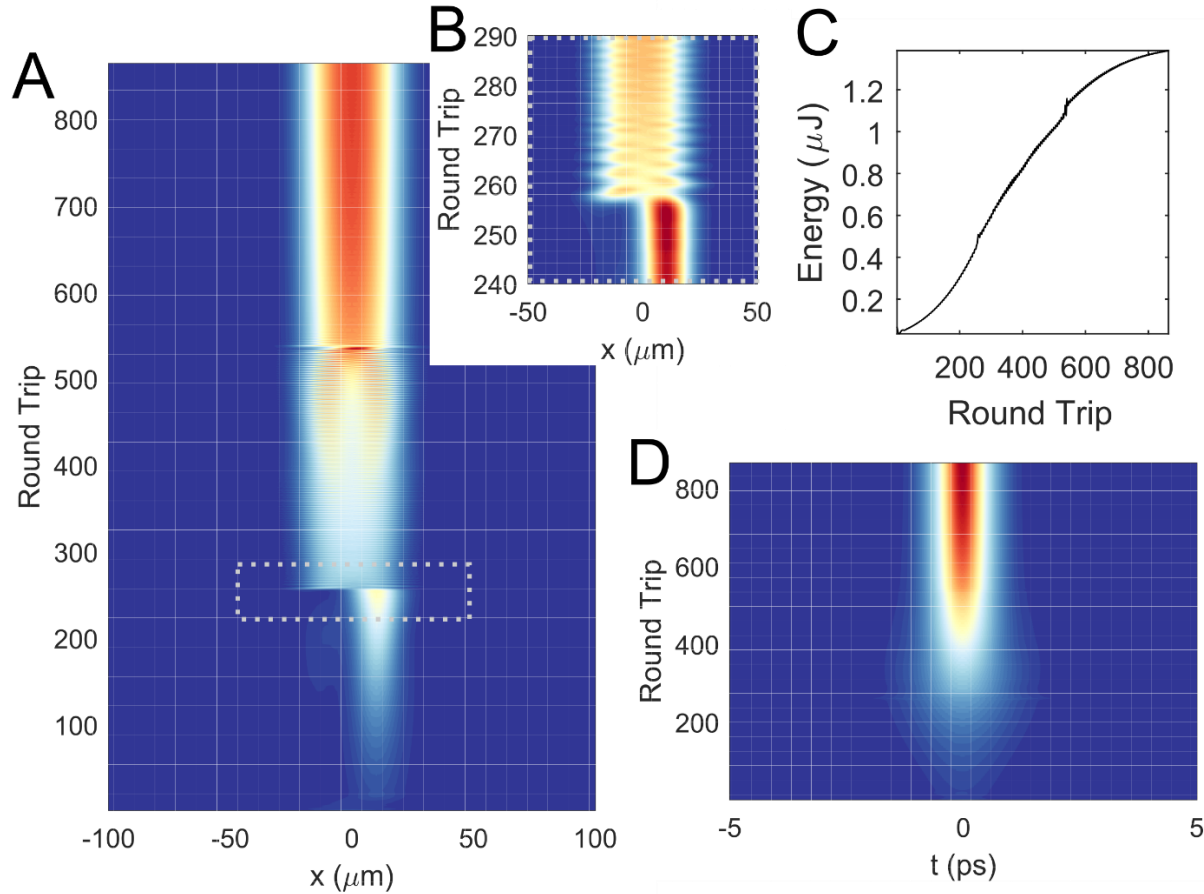

**Fig. S18: Numerical simulation showing different spatiotemporal mode-locking regimes in the second cavity.**
A simplified spatiotemporal GNLSE described previously models the system. A stable pulse is launched into the cavity and the saturation energy is increased by 5% every 6 round trips. **A**. The output spatial profile (**B**. zoom on a transition region). **C**. Energy as a function of round trip, and **D**. temporal profile for increasing round trip. The steady-state intracavity pulse evolution corresponding to fixed gain saturation energy at the highest level shown here is plotted in Fig. S19. For the results shown here, the cavity comprises a 50 cm gain fiber with a 1D index profile similar to the MM fiber used in experiments (Fig. S17). A 50 µm Gaussian (all bandwidths listed are full-width-at-half-maximum) spatial filter and 30 nm Gaussian spectral filter are applied to the spatiotemporal field envelope after saturable absorption and output coupling. The gain spectrum is modelled as a Gaussian with 40 nm bandwidth, and the small signal gain is taken to be 25 dB. The modulation depth of the saturable absorber is taken to be 100%, although stable pulses are observed for lower values. The output coupling ratio is taken to be 90%, and a lumped loss from fiber coupling of 50% is applied once per roundtrip. The dispersion of the fiber is assumed to be 20 fs$^2$/mm. Similar cavity parameters apply for subsequent simulations, with exceptional parameters noted.

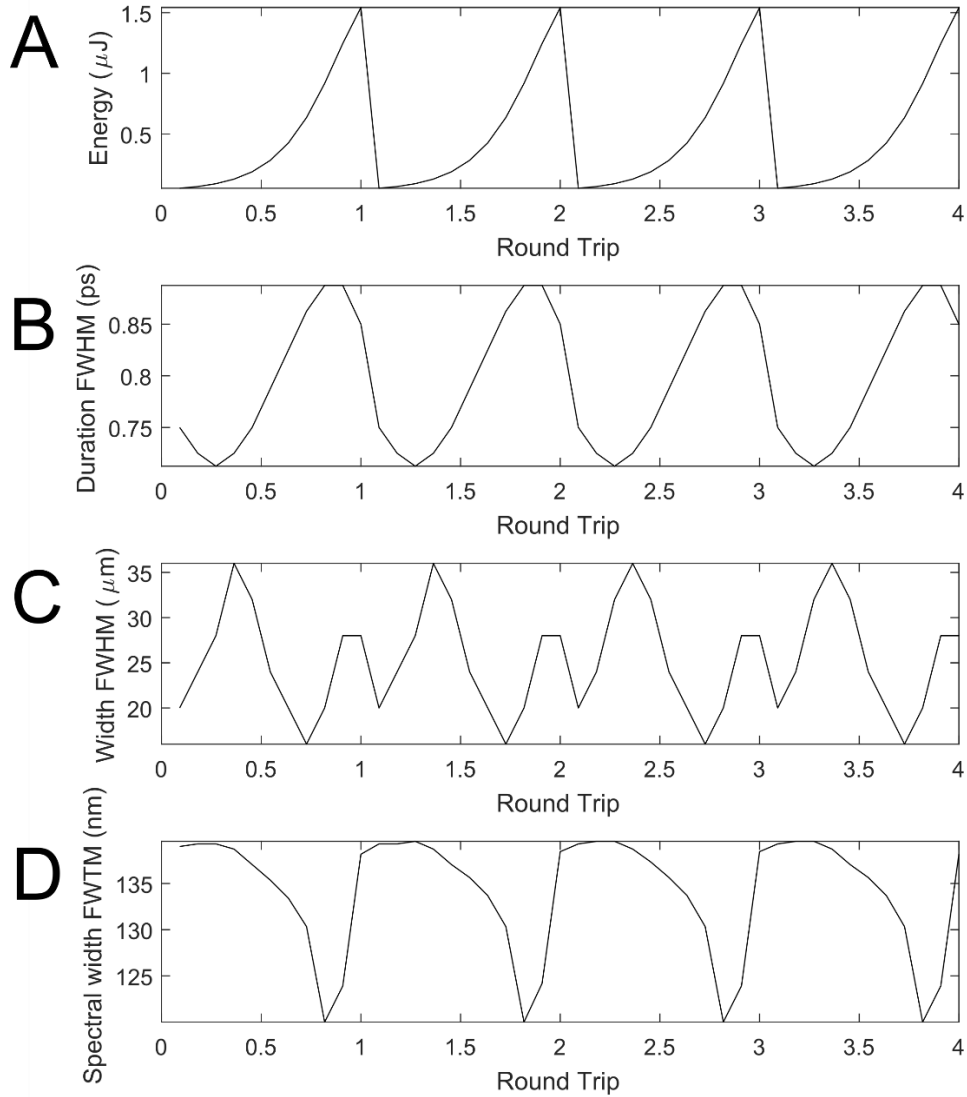

**Fig. S19: Steady-state intracavity pulse evolution corresponding to the highest-energy condition shown in the previous figure.**
The maximum gain saturation for the sequence shown in Fig. S18 is held constant. The figures show the evolution of **A**. the intracavity pulse energy, **B**. the intracavity pulse duration, full width at half maximum, **C**. the spatial width, and **D**. the spectral bandwidth, full width at 10% maximum. All parameters of the pulse exhibit a periodic evolution, and the spatial breathing shows evidence for the spatially-multimode nature of the pulse. The evolution continues indefinitely, but only the first 4 round trips are shown for clarity. These plots illustrate that, similar to the cavity where only Kerr effects are included, the spatiotemporally mode-locked pulse breathes in energy, duration, spatial dimensions and spectral bandwidth. All these quantities are periodically reduced by spectral and spatial filtering (and for the energy, output coupling and linear losses). Importantly, these suggest that mostly similar physics is responsible for spatiotemporal mode-locking in this kind of cavity. The strong spatial breathing shown in **C**. illustrates that, despite the stable Gaussian output profile of the pulse, the field within the cavity is multimode.

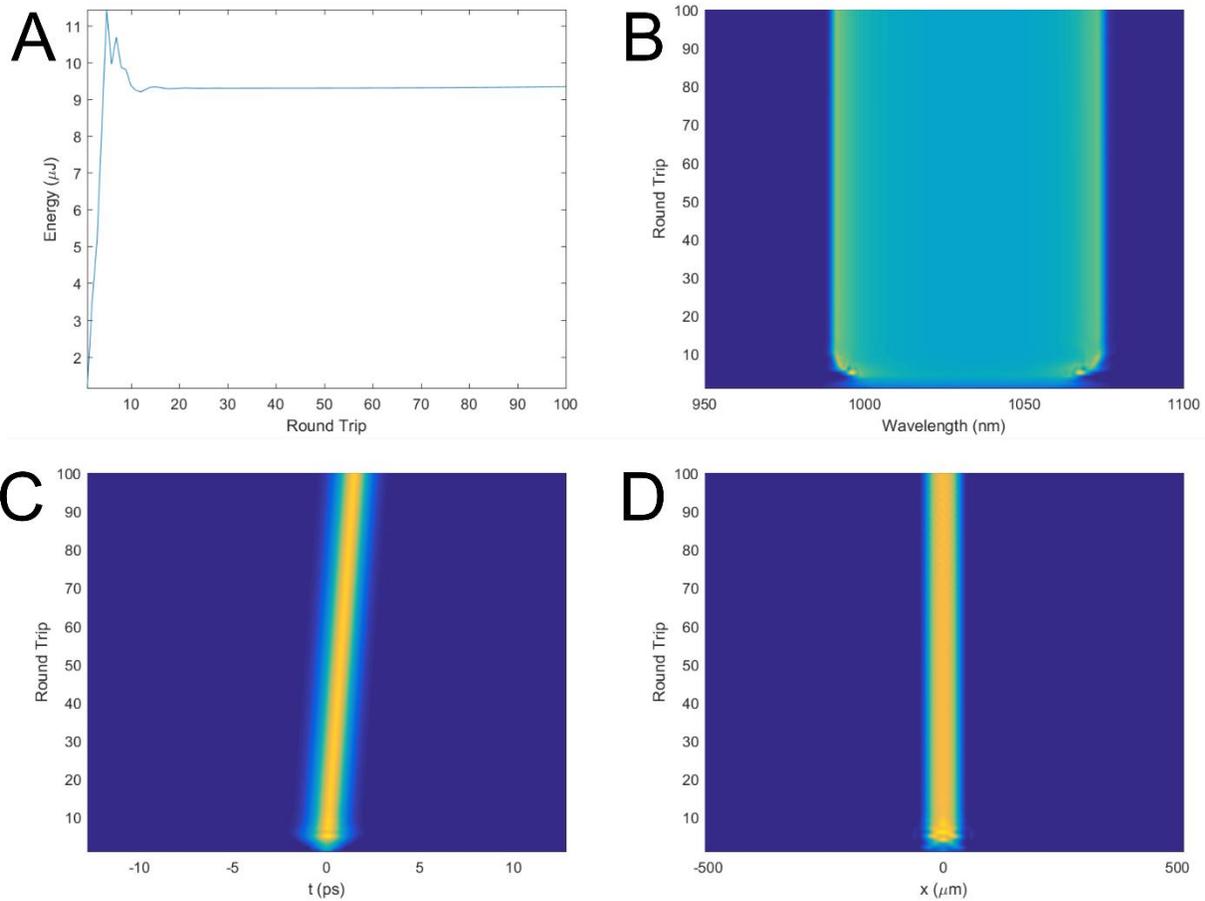

**Fig. S20: Numerical simulation of mode-locked pulse formation in a larger fiber multimode fiber cavity.**
The plots show **A**. the output energy, **B**. spectrum, **C**. time and **D**. spatial profile. To investigate scaling to higher performance, we consider a fiber twice the size (but similar numerical aperture and profile) as the one shown in Fig. S17. Here we include stimulated Raman scattering and self-steepening in the model, since these higher-order effects are expected to become important for limiting performance levels. The results show stable, nearly 10 µJ pulses can form. These pulses have a nearly Gaussian spatial profile, are identical from round-trip to round-trip, and have a slightly shifted group velocity.

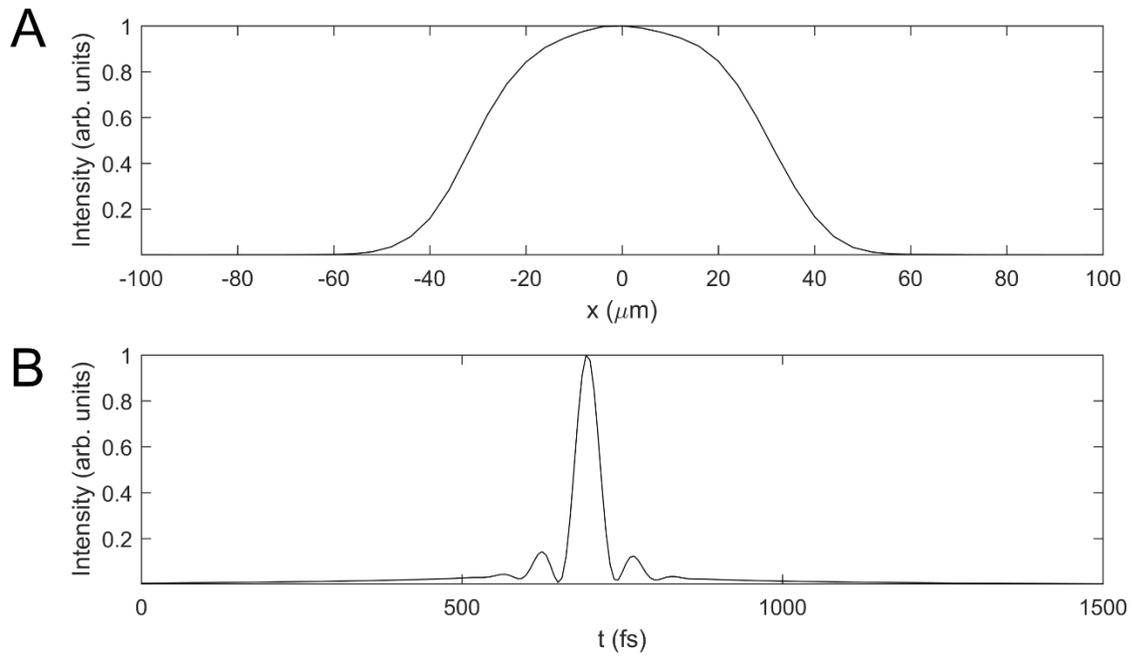

**Fig. S21: High energy pulse in numerical simulation of larger high-power multimode cavity.**
**A.** The stable spatial profile and **B**. the dechirped pulse temporal profile corresponding to the previous figure. The resulting peak power is roughly 0.2 GW.

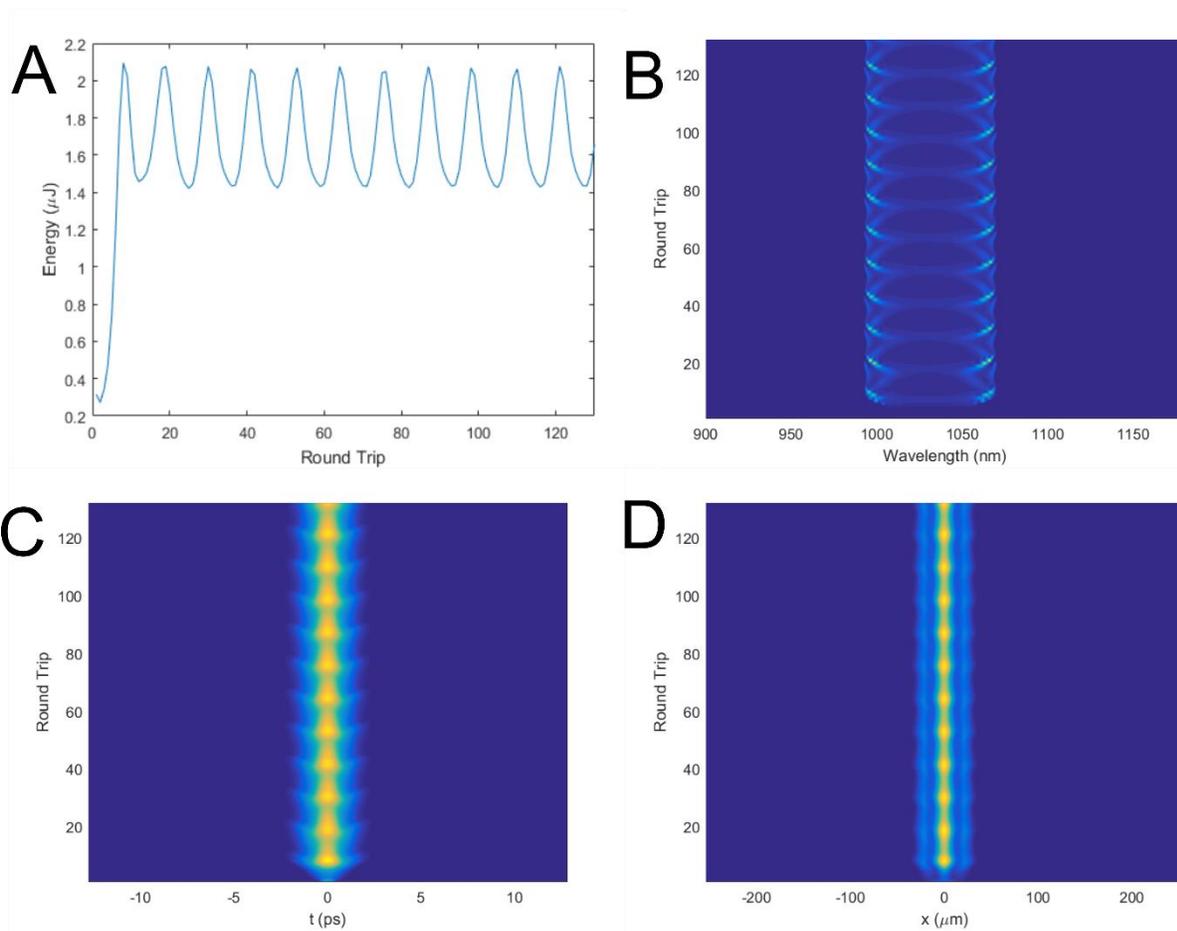

**Fig. S22: Numerical results in the high-power cavity showing periodic oscillations of a spatiotemporally-mode-locked pulse for spatial filtering below the fundamental mode size.** The plots show **A**. the output energy, **B**. spectrum, **C**. time and **D**. spatial profile. All evolve with a period of ~ 10 round trips. This behavior is observed for a spatial filter roughly ¼ the fundamental mode size.

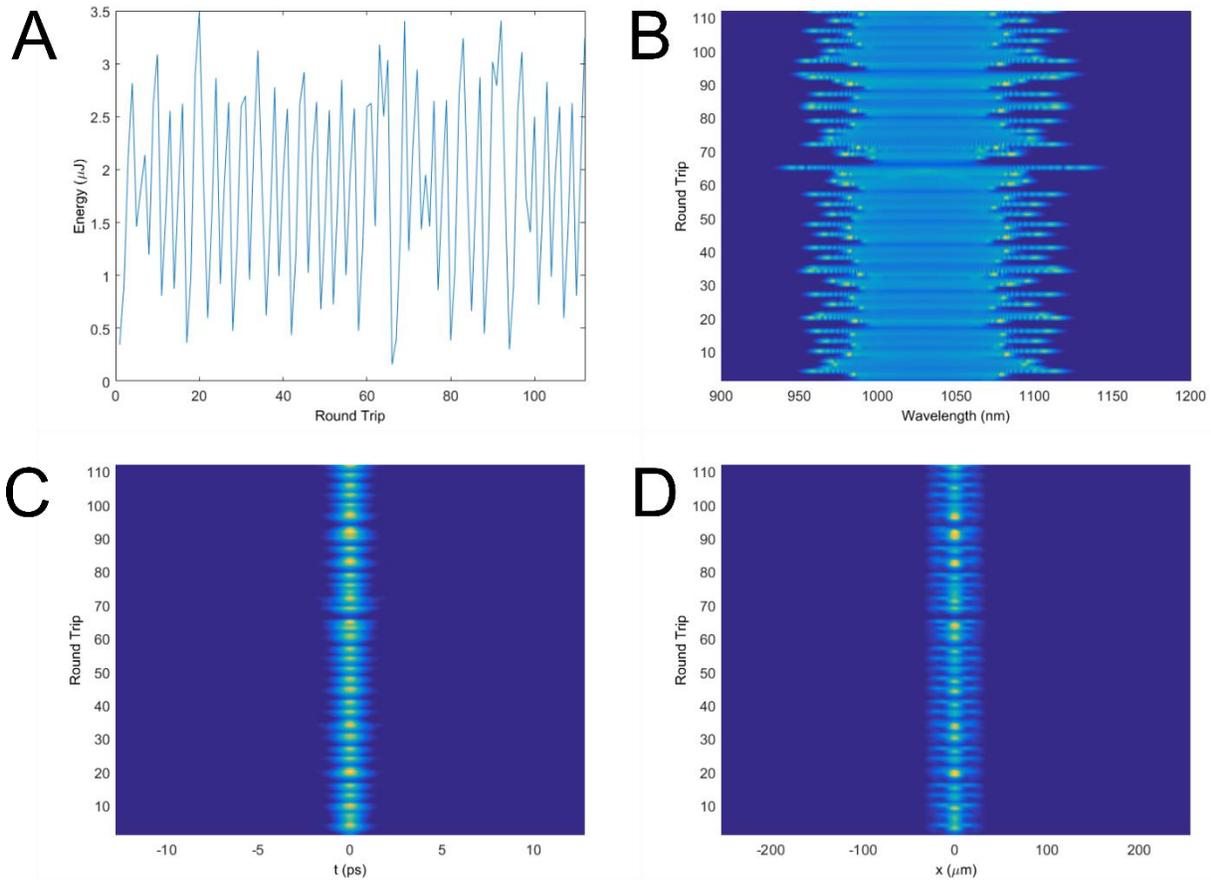

**Fig. S23: Numerical simulation of aperiodic evolution of a spatiotemporally-mode-locked pulse in the high-power cavity for strongly sub-fundamental spatial filtering.**
The plots show **A**. the output energy, **B**. spectrum, **C**. time and **D**. spatial profile. The dynamics show chaotic pulsations but without noise-like pulses. The observed behavior occurs for spatial filters <1/8 the fundamental mode size. Transform-limited peak powers near 0.25 GW occur despite the relatively small average mode-field diameter.

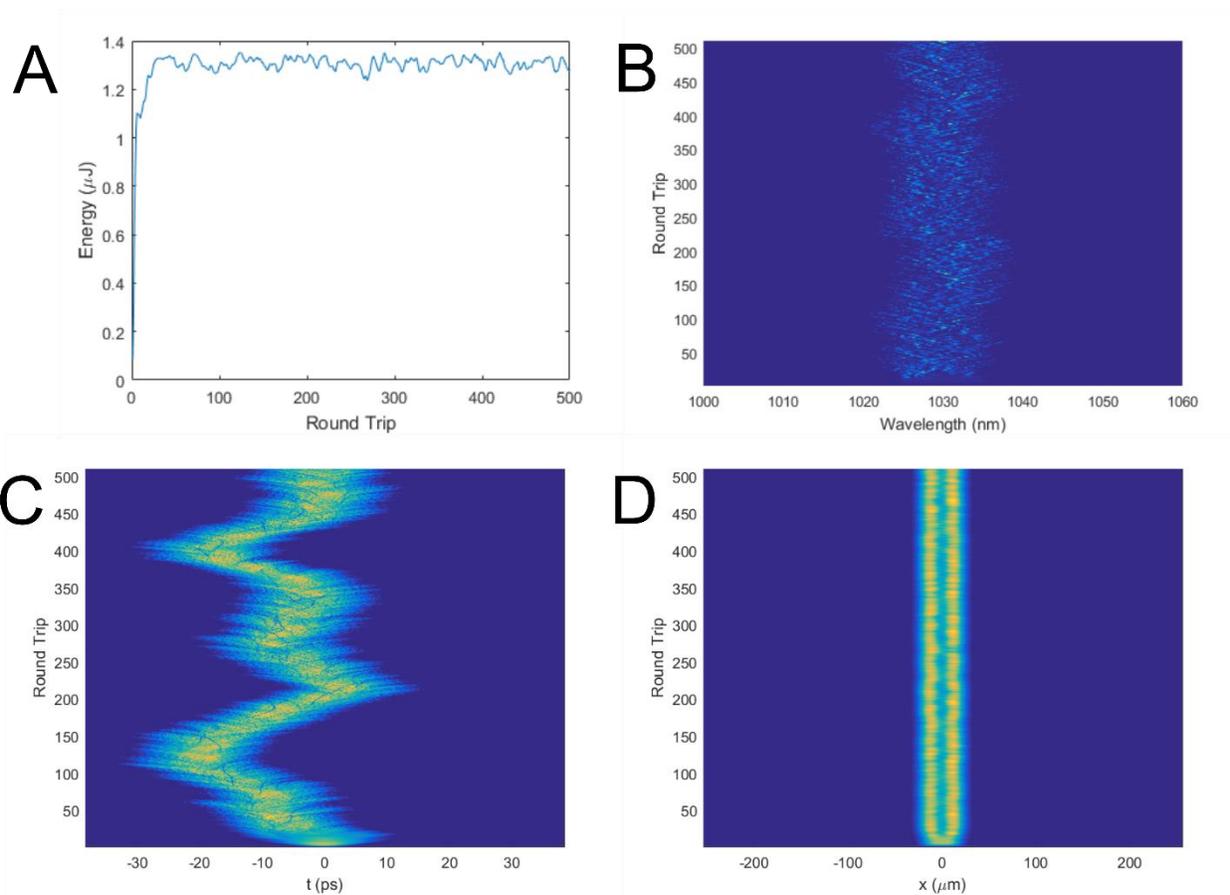

**Fig. S24: Example numerical result in the high-power cavity showing noise-like multimode pulse formation.**
The the plots show **A**. the output energy, **B**. spectrum, **C**. time and **D**. spatial profile. The energy is roughly stable, but the pulse exhibits a random-walk behavior that results from the complex changing mixture of modes that make it up. Dark pulse traces can be seen in the pulse over multiple round trips.

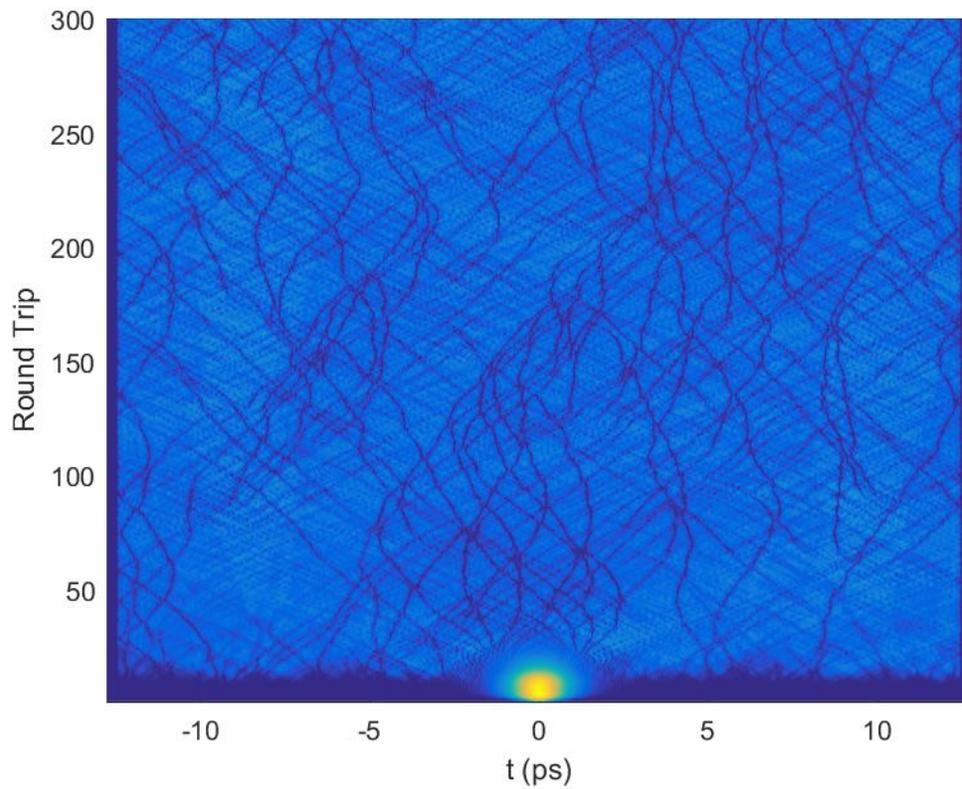

**Fig. S25: Example of numerical simulation showing multimode dark soliton formation.**
The other aspects of the evolution are similar to the previous figure.

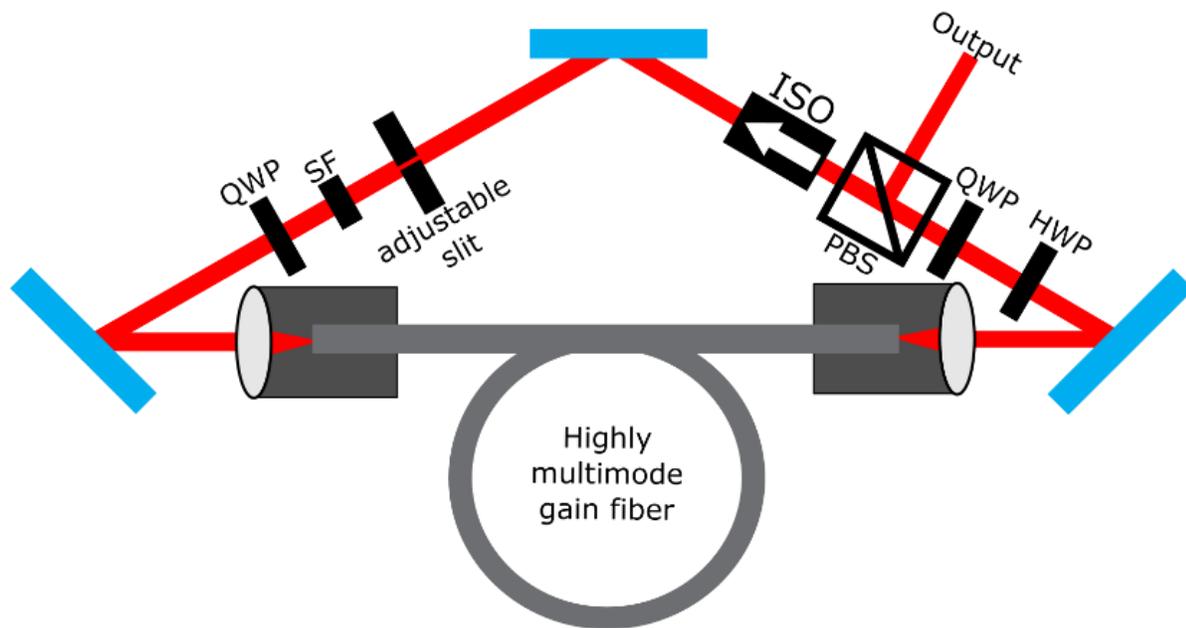

**Fig. S26: Schematic diagram of the cavity based entirely on a multimode gain fiber.** Nonlinear polarization rotation is used as an effective saturable absorber, controlled by the output coupling polarizing beamsplitter (PBS) and the half-wave and quarter-wave plates (HWP and QWP). The isolator ensures unidirectional light propagation in the ring cavity. An adjustable slit or pinhole provides spatial filtering, while a spectral filter (SF) is implemented using an interference filter.

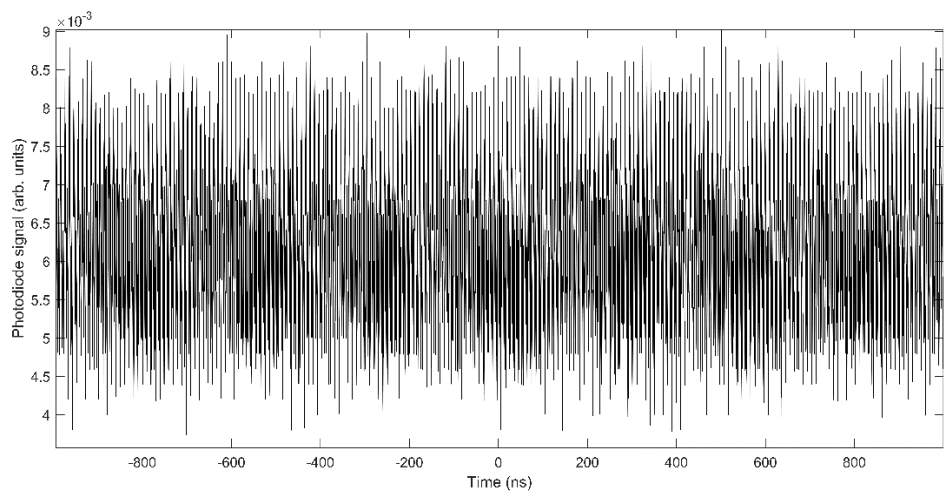

**Fig. S27: Photodetector measurement corresponding to pump power 18 W in Fig. 3b.**

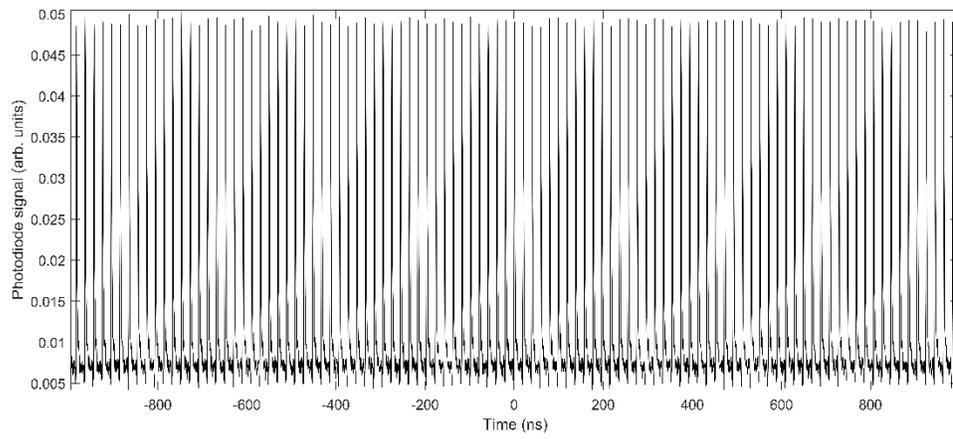

**Fig. S28: Photodetector measurement corresponding to pump power 30 W in Fig. 3b.**

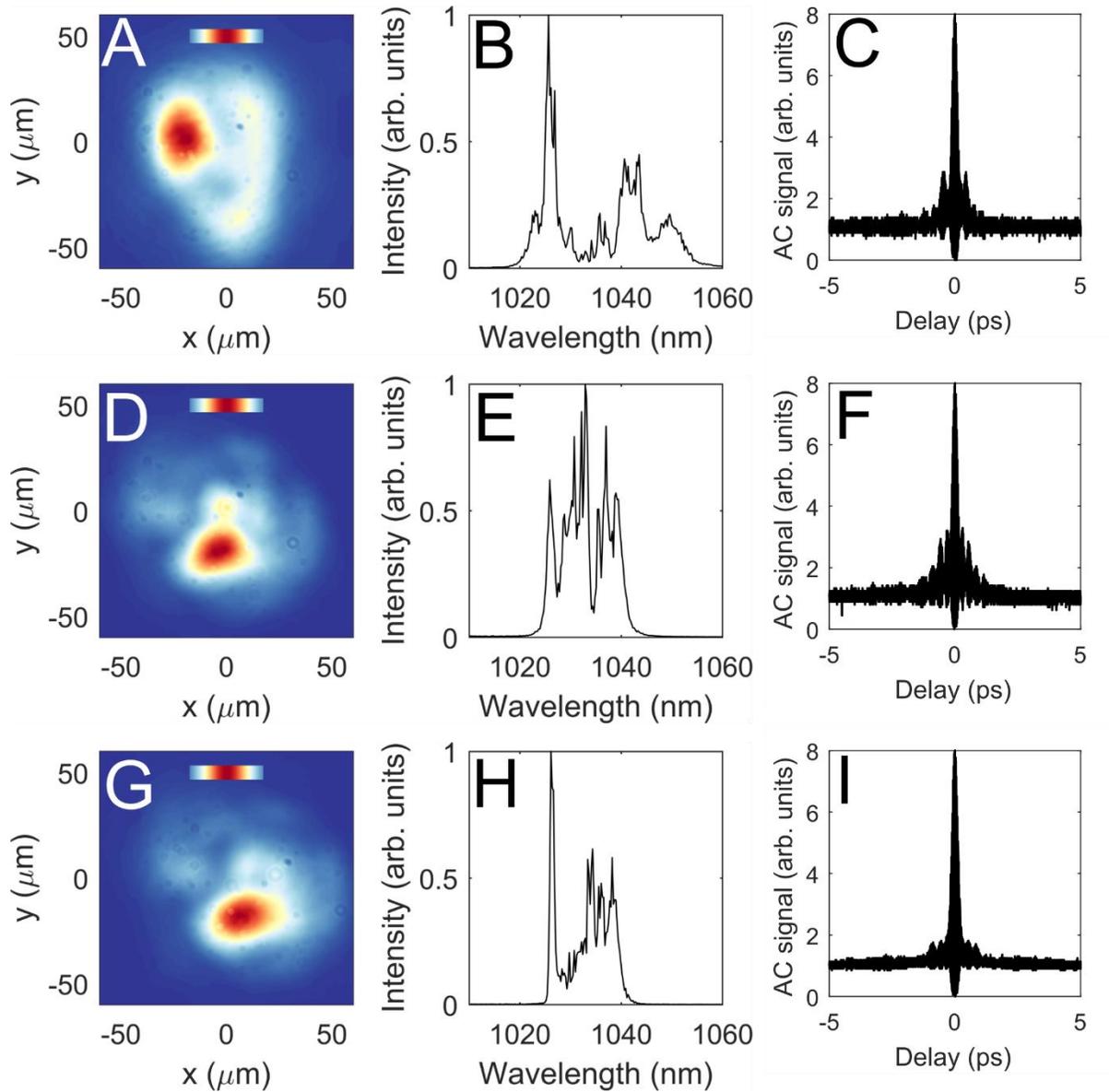

**Fig. S29: Examples of multimode mode-locked state observed experimentally.**
**A-C**, **D-F**, and **G-I** show the near-field beam profile, spectrum, and dechirped autocorrelation for different mode-locked states observed in the same cavity for different waveplate positions and mirror alignments. For **A-C**, the pulse energy is 100 nJ, the transform limited duration is 100 fs and the measured pulse width is 140 fs assuming a 1.4 deconvolution factor. For **D-F**, the pulse energy is 82 nJ, and the transform limited and measured durations are both 170 fs. For **G-I**, the pulse energy is 130 nJ and the transform limited and measured pulse durations are both 190 fs. The scale bar in **A,D,G** shows the Gaussian profile of the fundamental mode of the fiber.

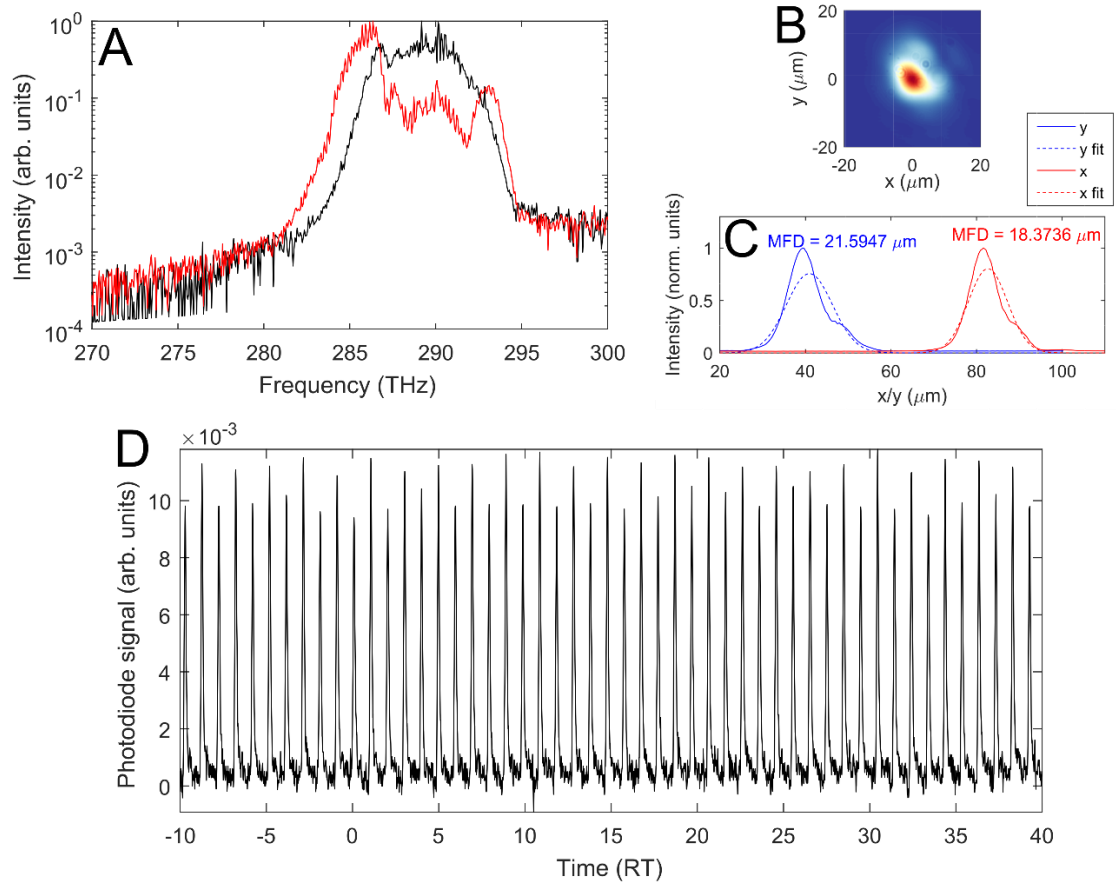

**Fig. S30: Assessment of peak power, example 1.**
**A**. The red curve shows the spectrum after 5-m of HI1060 fiber, for a coupled power of 4 mW. The black shows the spectrum measured using a long 400 μm diameter multimode fiber. In 5-m of HI1060 fiber, the spectral width at $10^{-1}$ ($10^{-2}$) broadens by a factor of 43% (20%) compared to the oscillator output spectrum. Comparison with the numerically-calculated spectral broadening for transform-limited Gaussian pulses in the same fiber implies a peak power of ~0.4-1.0 kW (uncertainty here stems from the uncertain initial pulse duration). For the condition shown, the oscillator's repetition rate is 62.5 MHz and the output power is 2.0 W (pulse energy 32 nJ). The beam profile (**B, C**) is multimode and the pulse train exhibits multi-round trip periodicity, shown below. The coupled sample of 4 mW corresponds to 64 pJ, which for the ~100 fs duration implies ~0.6 kW peak power. **C** shows Gaussian fit to cuts through maximum intensity. **D** shows the pulse train measurement. RT is the cavity round trip time. The pulse train exhibits oscillations over multiple round trips, which is a common but not typical feature of the multimode mode-locked states we observe.

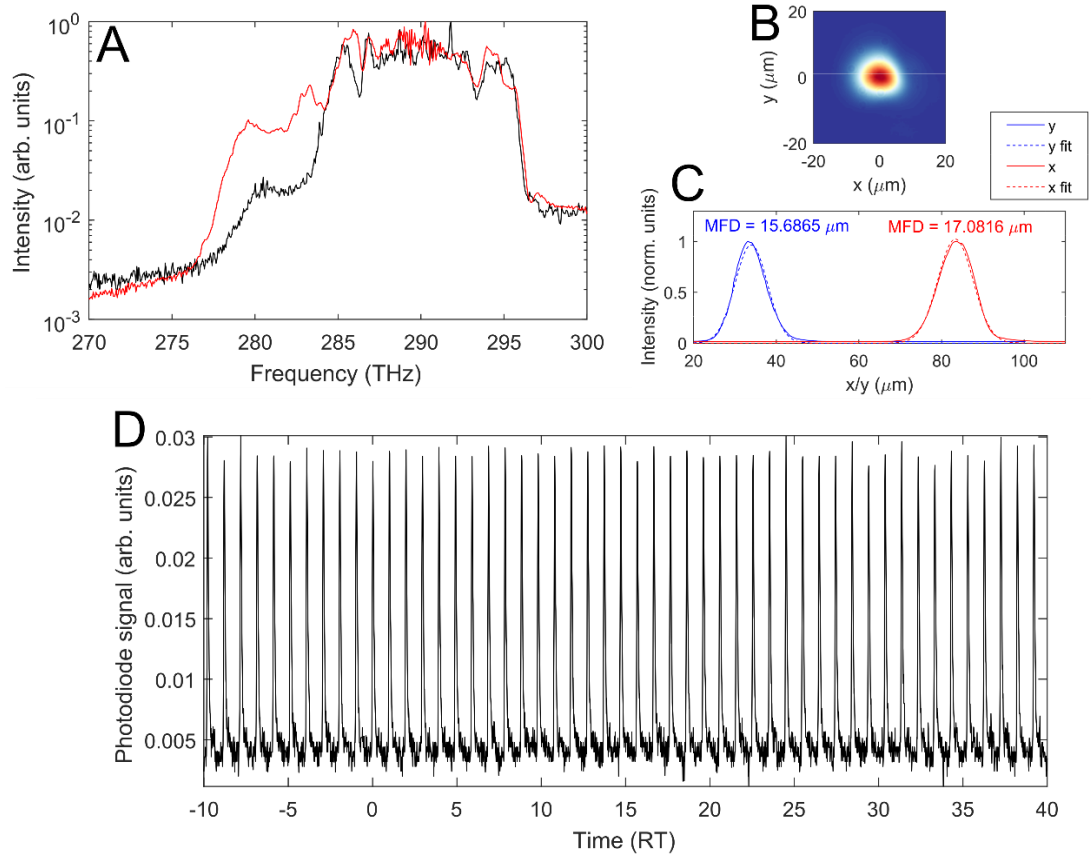

**Fig. S31: Assessment of peak power, example 2.**
**A.** For 5 mW coupled into 5-m of HI1060 fiber, the spectral width at $10^{-1}$ ($10^{-2}$) broadens by a factor of 33% (20%) compared to the measured spectrum in the large MMF. These correspond to ~0.6 to 1.0 kW peak power by comparison with numerical propagation of a transform-limited Gaussian pulse through the same fiber. **B.** Near-field beam profile corresponding to the mode-locked state considered in the previous figure. The beam is larger than the fundamental mode by >20%, but is evidently dominated by the fundamental mode. **C.** Gaussian fits to cuts through the maximum intensity. **D** shows the pulse train measurement. RT is the cavity round trip time.

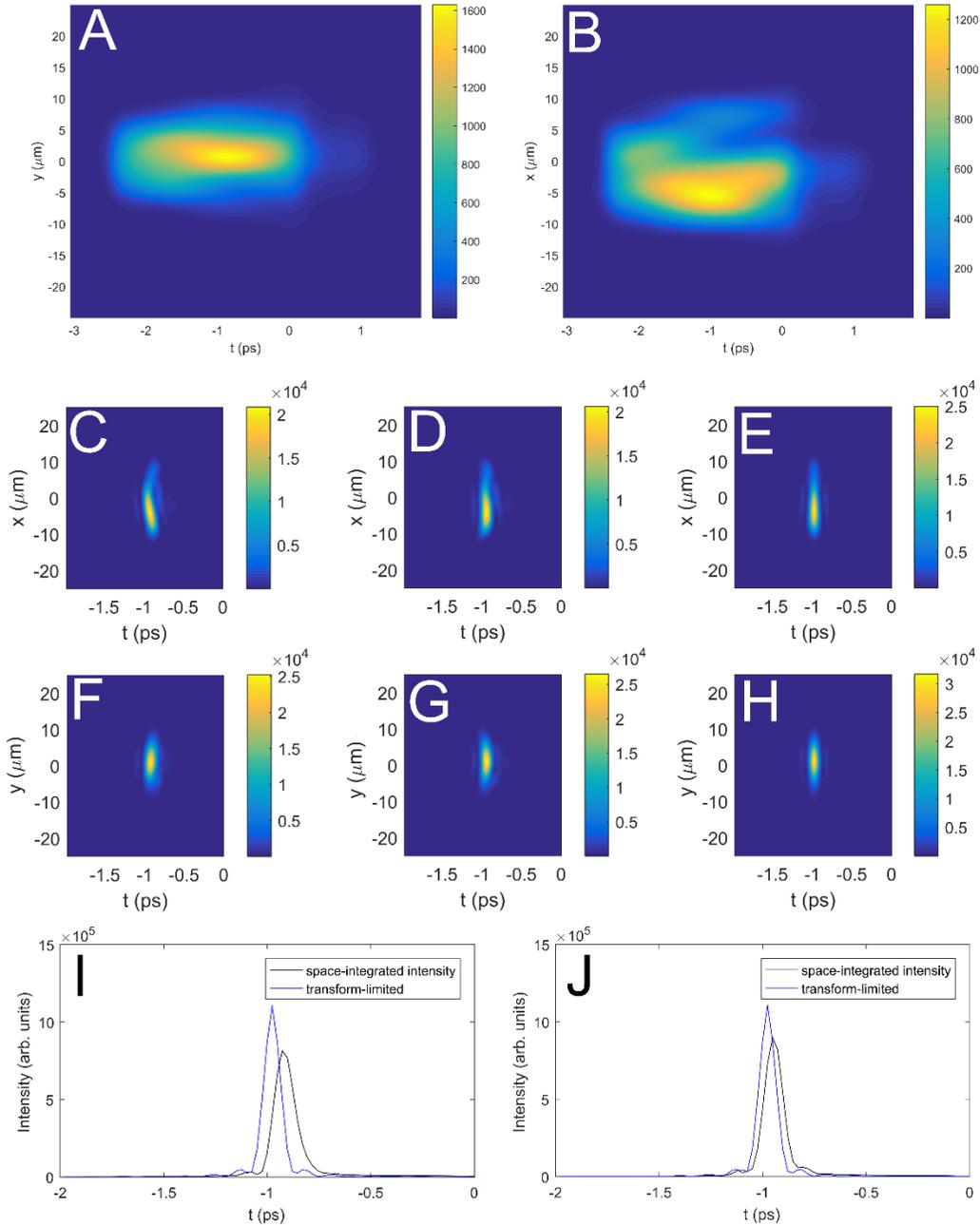

**Fig. S32: Spatiotemporal dispersion for the simulated pulse shown in Figs. S5-8.**
**A-B.** Spatiotemporal intensities for the output chirped pulse for numerical simulations of a MM mode-locked laser. **A.** the intensity integrated over the x-dimension, **B.** the intensity integrated over the y-dimension. **C.** the y-integrated spacetime profile after optimal group velocity dispersion compensation only, **D**. the optimal profile after spatiotemporal dispersion compensation, and **E.** the temporally-transform limited spatiotemporal intensity profile. **F-H** show the same for the x-integrated profile. Quadratic spatiotemporal dispersion means that we apply quadratic spectral phase (i.e., as introduced by an ideal grating compressor) and quadratic spatial group delay (i.e., such as is introduced by a glass lens). Here, the amount of quadratic spatial group delay applied is quite small, and is comparable to the amount introduced by the collimating and imaging lenses used in our experiments. **I-J.** The plots show the intensity profile of the pulse after integrating over both spatial dimensions. The blue curves show the temporally-transform limited pulse intensity, while the black curves show **I.** the intensity after group velocity dispersion compensation only, and **J.** after optimal quadratic spatiotemporal dispersion compensation.

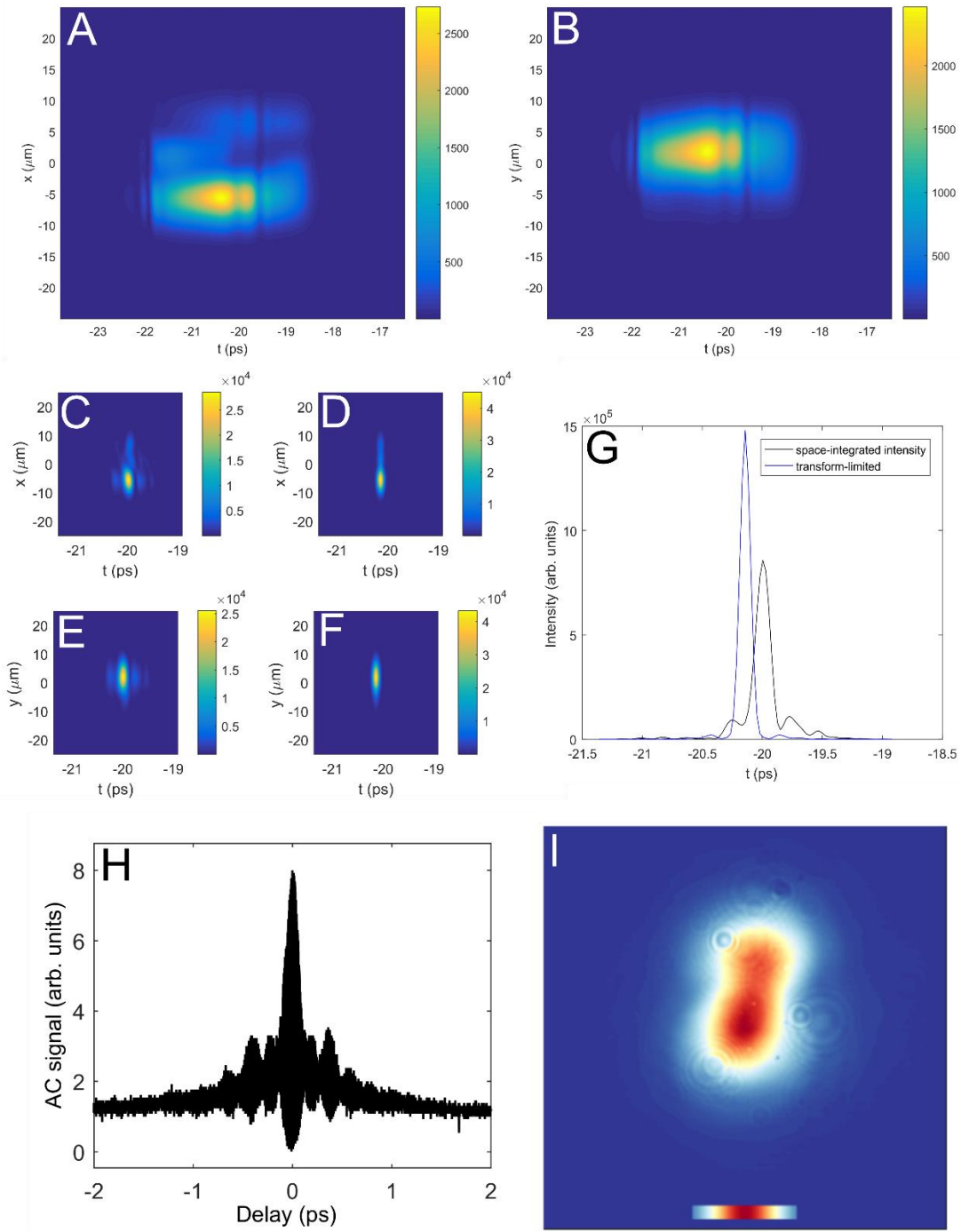

**Fig. S33: Spatiotemporal dispersion for the simulated pulse shown in Figs. S9-12.**
**A-B.** Spatiotemporal intensities for the output chirped pulse for numerical simulations of a MM mode-locked laser. **A.** the intensity integrated over the x-dimension, **B.** the intensity integrated over the y-dimension. **C-G.** Result of quadratic spatiotemporal dispersion compensation, compared to the transform-limited pulse. **C** and **E** show the spatiotemporal profiles, integrated over each spatial dimension, after applying the quadratic spatial and spectral phases that produce the maximum peak intensity. **D** and **F** show the corresponding transform-limited spatiotemporal profiles. **G** shows the temporal pulse intensity, integrated over both spatial dimensions, for the transform-limited and optimal quadratic spatiotemporal dispersion compensation. For this highly multimode pulse, mode-locking is only marginally stable and the optimally quadratically-compensated pulse has a peak intensity ~50% of the transform-limit, with prominent temporal sidelobes. **H-I.** Experimental measurements of a spatiotemporally mode-locked pulse exhibiting similar features as the previous numerical results. The results were obtained using a similar cavity as the first one described in the methods. **H.** the interferometric autocorrelation, and **I**. the near-field beam profile (the scale bar shows the Gaussian profile of the fundamental mode of the fiber, which has a 13 μm mode-field diameter). The comparison is not meant to be direct, since many features of the experiment cannot be adequately measured (e.g., the coupling coefficients into and out of the multimode GRIN fiber). However, the measurements suggest a similar spatiotemporal structure as depicted in **A-G**, suggesting that strongly spatiotemporally dispersed, multimode pulses are observed experimentally.

**Movie S1 (may be downloaded here:**
**http://science.sciencemag.org/highwire/filestream/700228/field_highwire_adjunct_files/1/aao0831s1.mp4 )**

This movie shows several demonstrations with the laser configuration shown in Fig. 2 and Figs. S1-4, and animated versions of Figs. S18-19. The demonstrations are of self-starting, environmental stability, the bistability observed at lower pump power, and the simulated starting dynamics.